%% file: main.tex
\documentclass[journal=nalefd,manuscript=article]{achemso}
\usepackage[T1]{fontenc}
\usepackage[utf8]{inputenc}
\usepackage{graphicx}
\usepackage{amsmath,amssymb}
\usepackage{booktabs}
\usepackage{multirow}
\usepackage{siunitx}
\usepackage{tabularx}
\usepackage{longtable}
\usepackage{array}
\usepackage{threeparttable}
\usepackage{dcolumn}
\usepackage{caption}
\usepackage{subcaption}
\usepackage{float}
\usepackage{enumitem}
\usepackage{hyperref}
\usepackage{dcolumn}
\newcolumntype{d}[1]{D{.}{.}{#1}}

\mciteErrorOnUnknownfalse

\newfloat{scheme}{htbp}{los}
\floatname{scheme}{Scheme}
\newfloat{graph}{htbp}{loh}
\floatname{graph}{Graph}

\usepackage[utf8]{inputenc}

\DeclareSIUnit{\angstrom}{Å}

\newif\ifcomments
\commentstrue
\ifcomments
  \newcommand{\EvL}[1]{\textcolor{orange}{#1}}
  \newcommand{\AK}[1]{\textcolor{magenta}{#1}}
\else
  \newcommand{\EvL}[1]{}
  \newcommand{\AK}[1]{}
\fi

\title{Giant Rashba Splitting and Enhanced Nonlinear Berry-Phase Responses in Sliding-Tunable vdW MXene Heterostructures}

\author{Ali Sufyan}
\affiliation[Lund-MP]{NanoLund and Division of Mathematical Physics, Department of Physics, Lund University, SE-221 00 Lund, Sweden}
\affiliation[Lund-WISE]{Wallenberg Initiative Materials Science for Sustainability, Department of Physics, Lund University, SE-221 00 Lund, Sweden}
\affiliation[LTU]{Applied Physics, Division of Materials Science, Department of Engineering Sciences and Mathematics, Lule\aa\ University of Technology, SE-971 87 Lule\aa, Sweden}

\author{J. Andreas Larsson}
\affiliation[LTU]{Applied Physics, Division of Materials Science, Department of Engineering Sciences and Mathematics, Lule\aa\ University of Technology, SE-971 87 Lule\aa, Sweden}
\affiliation[LTU-WISE]{Wallenberg Initiative Materials Science for Sustainability, Lule\aa\ University of Technology, SE-971 87 Lule\aa, Sweden}

\author{Andreas Kreisel}
\affiliation[Uppsala]{Department of Physics and Astronomy, Box 524, SE-751 20 Uppsala, Sweden}

\author{Erik van Loon}
\email{erik.van_loon@fysik.lu.se}
\affiliation[Lund-MP]{NanoLund and Division of Mathematical Physics, Department of Physics, Lund University, SE-221 00 Lund, Sweden}
\affiliation[Lund-WISE]{Wallenberg Initiative Materials Science for Sustainability, Department of Physics, Lund University, SE-221 00 Lund, Sweden}
\affiliation[LINXS]{LINXS Institute of advanced Neutron and X-ray Science (LINXS), Lund, Sweden}

\begin{document}

%%\twocolumn[

\begin{abstract}

Chalcogen-terminated van der Waals MXenes (M$_2$CX$_2$; M = Nb, Ta; X = S, Se) provide a robust platform for exploring strong spin--orbit coupling and proximity engineering. To probe their tunability and guide optimization of emergent properties, we systematically examine sister compounds and propose M$_2$CS$_2$/CrBr$_3$ heterostructures that break time-reversal symmetry via proximity exchange coupling, enabling combined intrinsic magnetic and mechanical control. 
First-principles calculations reveal Rashba splitting up to 2.53\,eV\,\AA{} and valley-contrasting spin polarization in monolayers. These features drive strong second-order nonlinear responses, with pristine bilayer Ta$_2$CS$_2$ reaching a shift current of $|\sigma|_{\max} \approx 5$ \AA{} mA/V$^2$ and Nb$_2$CS$_2$/CrBr$_3$ attaining $|D|_{\max} \approx 18.44$ \AA{}. In M$_2$CS$_2$/CrBr$_3$ heterostructures, the ferromagnetic substrate induces a magnetization-reversible proximity exchange field with valley-selective conduction-band renormalization ($\Delta_{\rm val} \approx 50$ meV). Crucially, interfacial geometry, controlled by stacking inversion and lateral sliding, acts as a mechanical knob that continuously tunes the exchange–SOC interplay and bandgap, driving an emergent quantum anomalous Hall phase in the bilayer. 

\vspace{0.5em}

\end{abstract}

%%\noindent\textbf{Keywords:} van der Waals MXenes; Rashba; spin--orbit coupling; Berry curvature dipole; shift current; quantum spin Hall; quantum anomalous Hall

%%\maketitle

%%\vspace{1em}
%%]

\vspace{1.5em}

%%\section{Introduction}

The discovery of graphene~\cite{novoselov2005two} started the field of two-dimensional van der Waal materials. Compared to bulk solids, two-dimensional materials offer the opportunity to efficiently perform environmental engineering: heterostructures~\cite{geim2013van,xu2013graphene,sun2017substrate} combining different van der Waals materials make it possible to use proximity effects to transfer properties from one layer to the other. In this way, it is possible to combine physical effects that are difficult to find in individual materials. The weak van der Waals bond between the layers means that this kind of heterostructures can be made without dramatically distorting the lattice structure of the individual layers.

The proximity effects can be understood as modifications of the Hamiltonian of a monolayer due to the environment. A dielectric~\cite{raja2017coulomb,van2023coulomb,9trc-9865,PhysRevResearch.3.013265} or metallic~\cite{yang2025engineering} environment screens the Coulomb engineering and thus allows for control of exciton binding energies~\cite{raja2017coulomb} and phase transitions~\cite{van2023coulomb,9trc-9865,PhysRevResearch.3.013265}. The single-particle Hamiltonian can be tuned using pseudodoping \cite{shao2019pseudodoping} or via the hybridization with metallic substrates~\cite{hall2019environmental,PhysRevMaterials.3.044003}, providing tools to control the stability of electronically-driven phase transitions.

Proximity effects in van der Waals materials can also be used to control the magnetism, either by substrate engineering a magnetic monolayer~\cite{liu2020high} or by using magnetic substrates to induces magnetic effects in otherwise non-magnetic van der Waals materials. An example of the latter is the spin-orbit and exchange coupling that can be proximity-induced in multilayer graphene~\cite{PhysRevLett.132.186401,Seiler_2025}.

Every new classes of two-dimensional (2D) materials opens up a multitude of opportunities for van der Waals (vdW) engineering, since the number of possible heterostructures increases rapidly with the number of monolayers. Starting from graphene, related hexagonal materials such as boron-nitride and phosphorene~\cite{PhysRevB.89.201408} entered the scene, followed by the transition metal dichalcogenides, transition metal halides, and MXenes\cite{vahidmohammadi2021world,sokol2019chemical}. MXenes constitute a broad class of 2D transition-metal carbides, nitrides, and carbonitrides, traditionally obtained by selective etching of the A-layer from layered MAX phases, yielding surface-terminated M$_{n+1}$X$_n$T$_x$\allowbreak{} sheets whose termination chemistry strongly governs their properties and applications\cite{vahidmohammadi2021world,sokol2019chemical}. Beyond this etched paradigm, a distinct subset comprises chalcogen-terminated, intrinsically multilayer vdW-MXenes (also known as transition-metal carbochalcogenides), where S/Se terminations are integral to the structure and neighboring M$_2$CCh$_2$ sheets are stacked via weak vdW forces, analogous to laminated vdW solids such as TMDs\cite{beckmann1970kristallstrukturen, boller1992quaternary, pang2020nb, loni2025two, ding2023chemical, helmer2024computational}. Within this family, Nb$_2$CS$_2$, Ta$_2$CS$_2$, and Nb$_2$CSe$_2$ are experimentally established layered compounds synthesized via solid-state, topochemical, or molten-salt routes, and reported in layered polytypes including $P\bar{3}m1$ (trigonal) and $R\bar{3}m$ (rhombohedral 3R) stackings\cite{dahlqvist2024chalcogen}, underscoring their natural propensity to form few-layer structures. Ta$_2$CSe$_2$, while predicted theoretically and alluded to experimentally with partial Se termination, provides a complementary system for exploration\cite{ding2023chemical,loni2025two,helmer2024computational}. Importantly, intercalation-assisted delamination and exfoliation have produced atomically thin sheets of Nb$_2$CS$_2$ and Ta$_2$CS$_2$, with single- and few-layer flakes directly resolved by microscopy, providing an experimental foundation for exploring not only the monolayer limit but also thickness-dependent evolution in bilayer/trilayer regimes, both theoretically and in vdW heterostructures\cite{majed2022transition}. Similar potential exists for their Se-terminated counterparts.
 
To address this, we combine first-principles calculations, tight-binding modeling, and symmetry analysis to map the spin textures and Berry-phase responses of M$_2$CX$_2$ (M = Ta, Nb; X = S, Se) from isolated monolayers to magnetic heterostructures. All four monolayers are non-centrosymmetric and show a pronounced 
Rashba splitting at $\Gamma$ together with valley-contrasting spin polarization at K/K$'$, concentrating Berry curvature into 
momentum-selective hot spots that drive sizable second-order 
responses: the shift current reaches $|\sigma|_{\max}\approx 5~\text{\AA\,mA/V}^2$, and the Berry-curvature dipole (BCD) reaches $|D|_{\max}\approx 18.44$~\AA. These magnitudes 
place the family among the strongest reported 2D platforms for junction-free bulk photovoltaics and zero-field nonlinear Hall 
transport.

Interfacing M$_2$CS$_2$ with the ferromagnetic insulator CrBr$_3$ breaks both inversion and time-reversal symmetries, generating a magnetization-reversible proximity exchange field and a valley-selective conduction-band renormalization. The interfacial geometry itself acts as a mechanical control knob: stacking inversion and lateral sliding continuously tune the exchange--SOC interplay and the band gap, and stabilize an emergent quantum anomalous Hall phase in the bilayer heterostructure. Together, these results establish chalcogen-terminated vdW MXenes as a tunable platform in which Rashba physics, proximity exchange, and Berry-phase responses can be jointly addressed by magnetic and mechanical means.

\begin{figure*}[t]
  \centering
  % try larger trims; units = left bottom right top
  \includegraphics[
    width=\textwidth,
    height=0.9\textheight,
    keepaspectratio,
    trim=0.0cm 0.35cm 1.5cm 0.0cm,
    clip
  ]{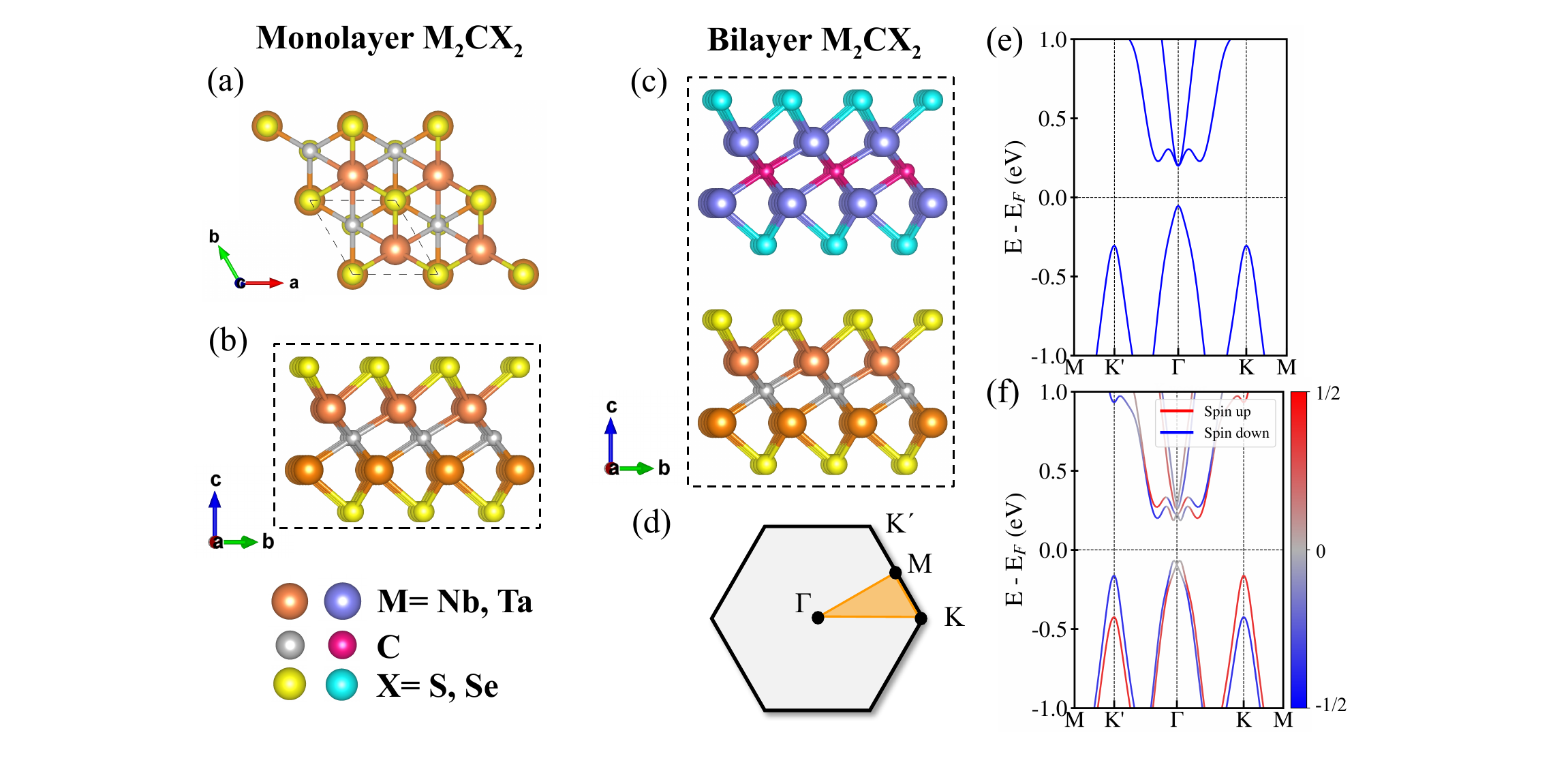}
  \caption{\textbf{Crystal structures and band structures of chalcogen-terminated vdW MXenes M$_2$CX$_2$ (M = Nb, Ta; X = S, Se).}
(a,b) Top and side views of the monolayer M$_2$CX$_2$ primitive cell.
(c) Side view of a representative bilayer M$_2$CX$_2$ structure.
(d) Two-dimensional hexagonal Brillouin zone and the high-symmetry path used for the band-structure calculations (M--K$'$--$\Gamma$--K--M).
(e) Electronic band structure of monolayer Ta$_2$CS$_2$ calculated without spin--orbit coupling (SOC).
(f) Electronic band structure of monolayer Ta$_2$CS$_2$ calculated with SOC, colored by the out-of-plane spin projection $\langle S_z \rangle$ (red/blue for $+1/2$/$-1/2$).}
  \label{fig:fig1}
\end{figure*}

Figure~\ref{fig:fig1} summarizes the structural models of the chalcogen-terminated MXenes investigated in this work, namely M$_2$CX$_2$ (M= Nb, Ta; X= S, Se). Figures~\ref{fig:fig1}a,b show the optimized top and side views of the lowest-energy monolayer geometry, which retains the characteristic M--C--M MXene backbone capped by chalcogen terminations on both surfaces in a hexagonal lattice\cite{helmer2024computational, lu2022enhancement, wang2022straintronic}. 

 To construct the bilayer (Fig.~\ref{fig:fig1}c), we stacked two monolayers in three high-symmetry configurations (AA, AB, and AC) and fully relaxed both atomic coordinates and the interlayer separation. In all cases, AB and AC initial configurations slide during relaxation and converge to the same AA-aligned structure, identifying AA stacking as the thermodynamic minimum. The corresponding hexagonal Brillouin zone and the high-symmetry path used throughout are shown in Figure~\ref{fig:fig1}d. Unless stated otherwise, we use Ta$_2$CS$_2$ as a representative platform for the main text, while the results for the remaining $\text{M}_2\text{CX}_2$ compounds are provided in the Supporting Information.

Figures~\ref{fig:fig1}e,f compare the electronic band structure of monolayer Ta$_2$CS$_2$ without and with spin--orbit coupling (SOC). Without SOC (Fig.~\ref{fig:fig1}e), the bands are spin-degenerate and the system is semiconducting. Upon adding SOC (Fig.~\ref{fig:fig1}f), the lack of inversion symmetry in the stable $P3m1$ monolayer lifts Kramers degeneracy, producing two characteristic spin splittings: (i) a Rashba-type splitting around the $\Gamma$ point, where the two branches acquire opposite spin character on opposite sides of $\Gamma$, and (ii) a pronounced valley-contrasting Zeeman-type splitting at the non-TRIM points $K$ and $K'$, with opposite spin polarization between the two valleys (spin--valley locking)\cite{sarmah2025rashba}.

One can qualitatively understand the Rashba-type splitting by considering $\mathbf k\cdot \mathbf p$ model including the symmetry of the d$_z$ like band-edge states at the $\Gamma$ point. The effective spin-orbit coupling originates from the interplay between inversion symmetry breaking, mixing with other orbitals of different $L$ and the magnitude of the atomic spin-orbit coupling. In this way, one can understand that the X=Se compounds with the heavier atom tend to have a stronger splitting compared to the X=S compounds, while the strongly increased mixing and reduction of the band gap in Ta$_2$CS$_2$ yields the largest overall band splitting, see Fig. S13 of the SM.
On the other hand, the splitting of the valence bands at the K and K$'$ points is due to the orbital character of $d_{x^2-y^2}\mp d_{xy}$ which is an eigenstate of $L_z=\mp 2$ and therefore the atomic spin-orbit coupling alone splits the bands in Zeeman-like eigenstates with constant splitting away from the high-symmetry points. The magnitude is governed by the atomic number of the transition metal. Finally, we observe a smaller splitting of spin-polarized conduction bands driven by a splitting of $L_z=\mp 1$ states yielding a splitting that is more than a factor 2 smaller. \\
Moreover, the two Ta atoms of the polar $P3m1$ monolayer occupy inequivalent Wyckoff positions on opposite ends of the intralayer S--Ta--C--Ta--S backbone, with Ta$_1$ sitting at lower $z$ and Ta$_2$ at higher $z$. Projecting the spin-resolved bands onto each Ta separately (Fig.~S16) shows that the two sublattices carry the band manifold unequally: Ta$_1$ dominates the entire valence-band region, including both the $d_{z^2}$-derived states near $\Gamma$ and the deeper $L_z=\mp 2$ ($d_{x^2-y^2}\mp i d_{xy}$) states responsible for the valley-Zeeman splitting at K/K$'$, whereas Ta$_2$ dominates the conduction-band region. This sublattice segregation is a direct consequence of the polar symmetry, which makes the two Ta sites inequivalent. The same sublattice segregation is found across the full M$_2$CX$_2$ family (Figs.~S14, S15, S17).

\begin{figure*}[t]
  \centering
  % try larger trims; units = left bottom right top
  \includegraphics[
    width=\textwidth,
    height=0.9\textheight,
    keepaspectratio,
    trim=0.0cm 0.35cm 1.5cm 0.0cm,
    clip
  ]{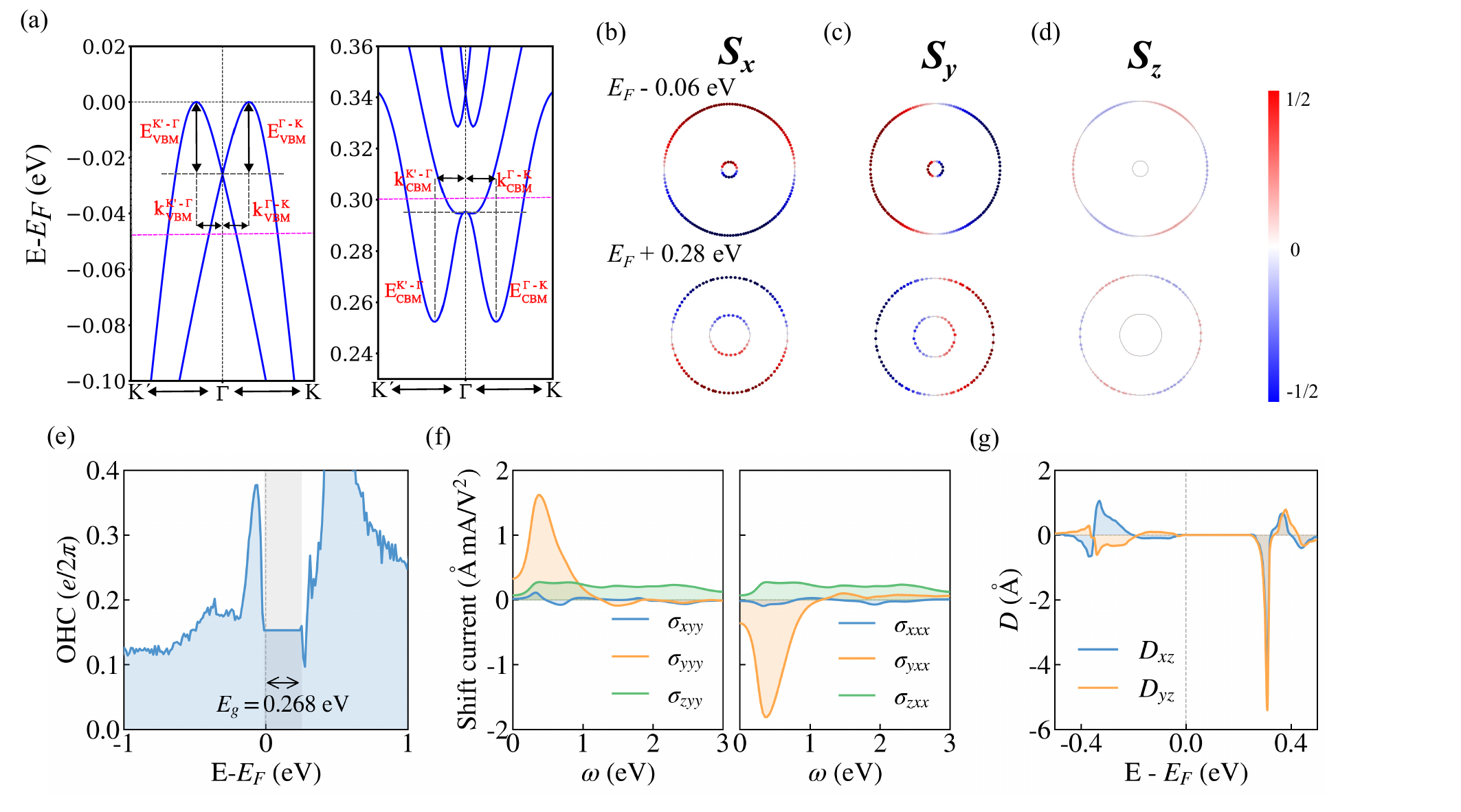}
  \caption{\textbf{Electronic structure, spin textures, and nonlinear response properties of monolayer Ta$_2$CS$_2$}. (a) Band dispersion along the $K$--$\Gamma$--$K$ path, with a zoomed view near $\Gamma$ spanning $\pm 0.2$ \AA$^{-1}$ along the selected high-symmetry path, highlighting the Rashba-type splitting. (b-d) Spin textures, shown through the $S_x$, $S_y$, and $S_z$ projections on constant-energy contours at $E_F-0.048$~eV and $E_F+0.30$~eV, revealing opposite chiralities of the Rashba-like spin winding. (e) Orbital Hall conductivity, $\mathrm{OHC}/(2\pi)$, as a function of $E-E_F$, showing a discontinuity across the band gap ($E_g=0.254$~eV). (f) Frequency-dependent shift-current tensor components $\sigma_{abc}(\omega)$. (g) Berry-curvature-dipole components as a function of $E-E_F$; the dashed line marks $E_F$.}
  \label{fig:fig2}
\end{figure*}

To quantify the SOC-driven Rashba splitting near $\Gamma$, we analyze the band edges along $K$--$\Gamma$--$K$ (Fig.~\ref{fig:fig2}a). SOC shifts the extrema away from $\Gamma$ and yields inner/outer branches characterized by $(k_R,E_R)$. From the VBM we obtain $k_R^{\mathrm{VBM}}=0.0622$~\AA$^{-1}$ and $E_R^{\mathrm{VBM}}=26.3$~meV, giving $\alpha_R^{\mathrm{VBM}}=2E_R/k_R=0.847$~eV\,\AA. For the CBM, $k_R^{\mathrm{CBM}}=0.0710$~\AA$^{-1}$ and $E_R^{\mathrm{CBM}}=46.1$~meV, yielding $\alpha_R^{\mathrm{CBM}}=1.298$~eV\,\AA. Hybrid-functional (HSE06) calculations yield a modest enhancement of the Rashba strength, giving $\alpha_R^{\mathrm{HSE}}=1.16$~eV,\AA\ at the VBM and $1.66$~eV,\AA\ at the CBM (Fig.~S23). Together, these results place Ta$_2$CS$_2$ among systems with pronounced Rashba coupling at both band edges, implying robust spin--momentum locking that is accessible under either electron or hole doping. The Rashba character is further corroborated by the helical spin textures with opposite chirality in Figs.~\ref{fig:fig2}b-d.

Figure~\ref{fig:fig2}e displays the energy-resolved intrinsic orbital Hall conductivity (OHC), $\sigma^{L_z}_{yx}$\cite{bernevig2005orbitronics,tanaka2008intrinsic}, for monolayer Ta$_2$CS$_2$. Despite the nonmagnetic ground state, inversion asymmetry and SOC generate a finite orbital Berry curvature. The OHC remains nearly constant within the bulk gap and exhibits strong energy dependence upon doping. The modest plateau at charge neutrality ($\sim 0.16\,e/2\pi$) reflects time-reversal symmetry-enforced partial cancellation and the mixed Ta-$d$ character of the Rashba-active band edges, which reduces the net $L_z$ polarization of the occupied manifold.

Figure~\ref{fig:fig2}f displays the shift-current response of monolayer Ta$_2$CS$_2$ under linearly polarized light\cite{ibanez2018ab}. The nonlinear photocurrent exhibits a pronounced low-energy peak: for $E\parallel y$, $\sigma_{yyy}$ reaches $1.616$~\AA,mA/V$^{2}$ ($\hbar\omega=0.39$~eV), while for $E\parallel x$, $\sigma_{yxx}$ peaks at $-1.809$~\AA,mA/V$^{2}$ ($\hbar\omega=0.36$~eV). Other components remain below 0.3 \AA~mA/V$^2$, underscoring the strong in-plane anisotropy of the bulk photovoltaic response in this inversion-asymmetric system.

Figure~\ref{fig:fig2}g reports the total Berry-curvature dipole (BCD) components $D_{xz}$ and $D_{yz}$ of monolayer Ta$_2$CS$_2$, shown as a function of the Fermi-level shift\cite{sodemann2015quantum}. As expected for an insulator, the BCD vanishes throughout the bulk gap, but it is sharply activated once $E_F$ enters the band edges, where SOC-split Rashba/valley pockets sample Berry-curvature hot spots. In particular, electron doping produces a pronounced enhancement, with extrema of $D_{xz}=-4.41$~\AA\ and $D_{yz}=-5.51$~\AA\ at $E_F-E_F^{(0)}=0.31$~eV, underscoring a strong and highly tunable nonlinear Hall propensity near the conduction-band onset.

\begin{table*}[t]
\caption{\textbf{Band gaps, Z$_2$ numbers, and Rashba coefficients extracted near the VBM and CBM}. The Rashba coefficient is reported as $\alpha^{\mathrm{ext}} = 2E_R/k_R$ (magnitude shown). $|D|_{\max}$ is the maximum absolute Berry-curvature dipole (largest among the reported $D_{xz}$ and $D_{yz}$ values in the scanned $E_F$ window). $|\sigma|_{\max}$ is the maximum absolute shift-current response (largest among the reported tensor elements within $\omega \in [0,3]$ eV).}
\label{tab:rashba_gap_bcd_shift}
\centering

\setlength{\tabcolsep}{3.5pt}
\renewcommand{\arraystretch}{1.10}
\footnotesize

\sisetup{
  table-number-alignment = center,
  round-mode = places,
  round-precision = 2
}

\begin{tabular}{ll
  S[table-format=3.1,round-precision=1]
  S[table-format=3.1,round-precision=1]
  c
  S[table-format=1.2]
  S[table-format=1.2]
  S[table-format=1.2]
  S[table-format=1.2]
  S[table-format=2.2]
  S[table-format=1.2]
}
\toprule
& & \multicolumn{2}{c}{Band gap} & & \multicolumn{4}{c}{Rashba coefficient} & & \\
\cmidrule(lr){3-4}\cmidrule(lr){6-9}
Material & Thickness
& {$E_g$} & {$E_g^{\mathrm{HSE06}}$}
& {$Z_2$}
& {$|\alpha^{\mathrm{ext}}_{\mathrm{VBM}}|$} & {$|\alpha^{\mathrm{ext}}_{\mathrm{CBM}}|$}
& {$|\alpha^{\mathrm{ext,HSE}}_{\mathrm{VBM}}|$} & {$|\alpha^{\mathrm{ext,HSE}}_{\mathrm{CBM}}|$}
& {$|D|_{\max}$} & {$|\sigma|_{\max}$} \\
& &
{(meV)} & {(meV)} &
& {(\si{\electronvolt\angstrom})} & {(\si{\electronvolt\angstrom})}
& {(\si{\electronvolt\angstrom})} & {(\si{\electronvolt\angstrom})}
& {(\si{\angstrom})} & {(\si{\angstrom\milli\ampere\per\volt\squared})} \\
\midrule

\multirow{2}{*}{Nb$_2$CS$_2$}
& Monolayer & 171.8 & 190.0 & 0 & 0.31 & 0.61 & 0.44   & 1.04   & 10.93 & 0.65 \\
& Bilayer   & 7.0   & 7.8   & 1 & 0.50 & 0.69 & 0.90   & 1.18   & 6.88  & 0.87 \\

\multirow{2}{*}{Nb$_2$CSe$_2$}
& Monolayer & 270.2 & 330.0 & 0 & 0.22 & 0.13 & 0.30   & 0.09   & 0.49  & 0.44 \\
& Bilayer   & 70.8  & 75.0  & 0 & 0.32 & 0.06 & 0.21   & 0.14   & 0.15  & 0.87 \\

\multirow{2}{*}{Ta$_2$CS$_2$}
& Monolayer & 254.0 & 280.7 & 0 & 0.85 & 1.30 & 1.16   & 1.66   & 5.41  & 1.81 \\
& Bilayer   & 5.5   & 6.3   & 1 & 1.30 & 0.88 & 1.66   & 2.53   & 7.43  & 4.99 \\

\multirow{2}{*}{Ta$_2$CSe$_2$}
& Monolayer & 243.9 & 328.7 & 0 & 0.04 & 1.02 & 0.33   & 1.29   & 0.45  & 0.62 \\
& Bilayer   & 25.4  & 68.3  & 0 & 0.07 & 0.96 & 0.16   & 0.91   & 0.74  & 0.98 \\
\bottomrule
\end{tabular}
\end{table*}

\begin{figure*}[t]
  \centering
  % try larger trims; units = left bottom right top
%   \includegraphics[
%     width=\textwidth,
%     height=0.9\textheight,
%     keepaspectratio,
%     trim=1.2cm 0.0cm 1.7cm 0.2cm,
%     clip
%   ]{fig3.pdf}
%  trimming can be done on the figure file, setting height and width with keepaspectratio is missleading
    \includegraphics[width=\textwidth]{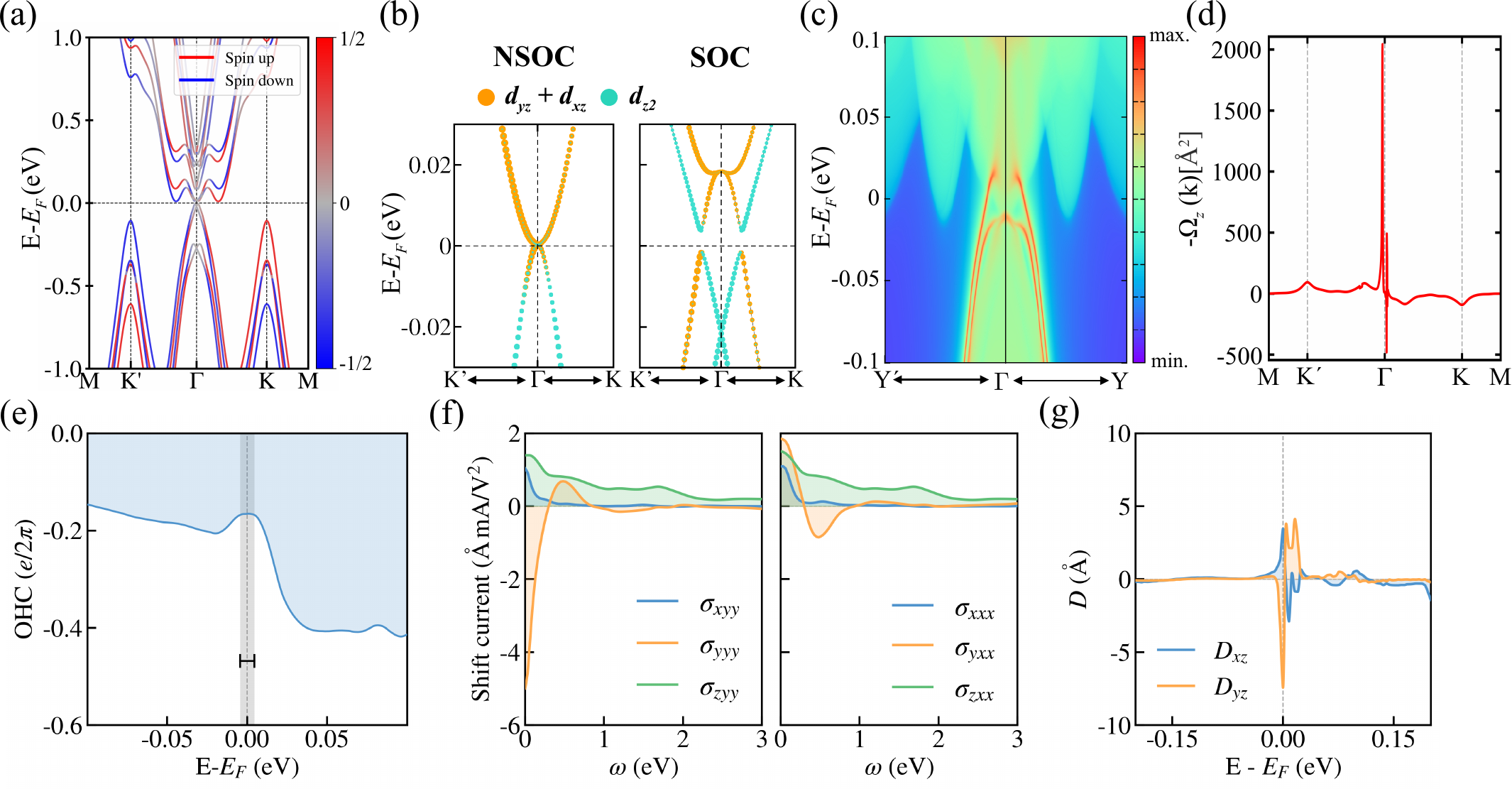}
  \caption{\textbf{Quantum spin Hall state and Berry-phase responses in bilayer Ta$_2$CS$_2$}. (a) SOC band structure along $M$--$K'$--$\Gamma$--$K$--$M$ with bands colored by the spin projection (color scale $\pm 1/2$). (b) Orbital-resolved bands near $\Gamma$ without SOC and with SOC, highlighting the SOC-induced band inversion between Ta-$d$ manifolds. (c) Edge spectral function/band structure of a semi-infinite ribbon, showing gapless helical edge modes connecting valence and conduction continua. (d) Berry curvature $\Omega_z(\mathbf{k})$ (integrated along the chosen $k$ path) exhibiting the characteristic SOC-driven redistribution consistent with a nontrivial $Z_2$ topology. (e) Energy-dependent orbital Hall conductivity $\sigma^{L_z}_{yx}$ with the bulk gap $E_g$ indicated. (f) Frequency-dependent shift-current tensor components $\sigma_{abc}(\omega)$. (g) Berry-curvature-dipole components $D_{xz}$ and $D_{yz}$ as a function of $E-E_F$.}
  \label{fig:fig3}
\end{figure*}

%%\subsection{Bilayer}
Figure~\ref{fig:fig3} illustrates the electronic and topological fingerprints of bilayer Ta$2$CS$2$. The SOC band structure (Fig.~\ref{fig:fig3}a) exhibits Rashba-type splitting at both band edges, with parameters $\alpha_R = 1.30$~eV\AA\ (VBM) and $0.886$~eV\AA\ (CBM). Orbital-resolved dispersions (Figure~\ref{fig:fig3}b) reveal that SOC opens a 5.5~meV gap at $\Gamma$ via Ta-$d$ band inversion between $(d_{xz}+d_{yz})$ and $d_{z^2}$ manifolds. Consistent with this SOC-gapped inversion, the $Z_2$ index is nontrivial ($Z_2=1$), identifying bilayer Ta$_2$CS$_2$ as a quantum spin Hall insulator. The ribbon spectrum (Fig.~\ref{fig:fig3}c) confirms gap-spanning helical edge states, though the narrow bulk gap leads to overlap with projected states. The Berry-curvature map (Figure~\ref{fig:fig3}d) reveals a sharp $\Omega_z(\mathbf{k})$ hot spot at $\Gamma$ and antisymmetric valley contributions at K/K$'$, reflecting strong, momentum-selective SOC hybridization.

Figure~\ref{fig:fig3}e shows the OHC $\sigma^{L_z}_{yx}$ of bilayer Ta$_2$CS$_2$. Within the gap, the OHC forms a plateau ($\sim -0.18\,e/2\pi$) similar to the monolayer. Figure~\ref{fig:fig3}f reveals a markedly enhanced low-energy shift-current response in the bilayer. For $E\parallel y$, $|\sigma_{yyy}|$ reaches $\approx 4.99$~\AA\,mA/V$^{2}$ as $\hbar\omega \to 0$, a threefold increase over the monolayer. A strong in-plane anisotropy persists, with $\sigma_{yxx} \approx 1.85$~\AA\,mA/V$^{2}$ at the low-energy onset. This near-gap enhancement stems from the SOC-reconstructed band edges, which amplify interband Berry-connection and shift-vector contributions. In terms of magnitude, the bilayer peak $|\sigma_{yyy}|\approx 4.99$~\AA\,mA/V$^{2}$ places Ta$_2$CS$_2$ among the highest-performing 2D shift-current materials, exceeding representative values reported for CrI$_3$ ($\sim$0.2)\cite{zhang2019switchable}, GeS ($\sim$0.1)\cite{rangel2017large}, WS$_2$ ($\sim$1)\cite{wang2017first}, Sb ($\sim$0.67)\cite{qian2023shift}, As ($\sim$0.57)\cite{qian2023shift}, Bi ($\sim$2)\cite{qian2023shift}, and Ti$_4$C$_3$ ($\sim$0.15)\cite{sufyan2025evidence} as well as the bulk of the responses found in large-scale screenings (e.g., the 326-material dataset in Ref. \cite{sauer2023shift}) in the same unit convention.

Figure~\ref{fig:fig3}g displays the BCD components $D_{xz}$ and $D_{yz}$ for bilayer Ta$_2$CS$_2$. Due to the narrow SOC gap, the BCD activates immediately upon $E_F$ shifting, yielding sharp, sign-changing features. The peak response occurs at charge neutrality, with $D_{xz}=3.46$~\AA\ and $D_{yz}=-7.43$~\AA, indicating an exceptionally tunable nonlinear Hall propensity. These magnitudes place Ta$_2$CS$_2$ at the upper end of reported 2D BCD responses, exceeding benchmarks such as mono/few-layer $T_d$-WTe$_2$ ($\sim$0.1--0.7~\AA)\cite{he2021giant}, MoS$_2$ ($\sim$0.4~\AA)\cite{son2019strain}, WSe$_2$ ($\sim$3~\AA)\cite{hu2022nonlinear}, Bi(110) ($\sim$1~\AA)\cite{jin2021enhanced}, and SnTe ($\sim$0.47~\AA)\cite{kim2019prediction}.

Table~\ref{tab:rashba_gap_bcd_shift} summarizes the near-edge metrics for all four MXenes. Two robust trends emerge: (i) Rashba coupling is consistently stronger in Ta-based compounds than Nb analogues, especially at the CBM, reflecting the larger Ta-derived SOC; (ii) thinning from monolayer to bilayer suppresses the band gap. HSE06 calculations corroborate these trends, supporting our $U_{\mathrm{eff}}$ choice. These band-edge characteristics correlate with nonlinear responses: bilayer Ta$_2$CS$_2$ exhibits the largest shift currents, whereas selenides show weaker magnitudes. Interestingly, BCD depends on both Rashba strength and Berry-curvature hot spots; monolayer Nb$_2$CS$_2$ attains the largest $|D|_{\max}$ ($\sim$10.93~\AA). Like the Ta-analogue, bilayer Nb$_2$CS$_2$ is a quantum spin Hall insulator ($Z_2=1$; see Fig.~S5). Additional data for other family members are provided in Figs.~S1--S4.

\begin{figure*}[t]
  \centering
  % try larger trims; units = left bottom right top
  \includegraphics[
    width=\textwidth,
    height=0.9\textheight,
    keepaspectratio,
    trim=0.2cm 0.4cm 0.2cm 0.4cm,
    clip
  ]{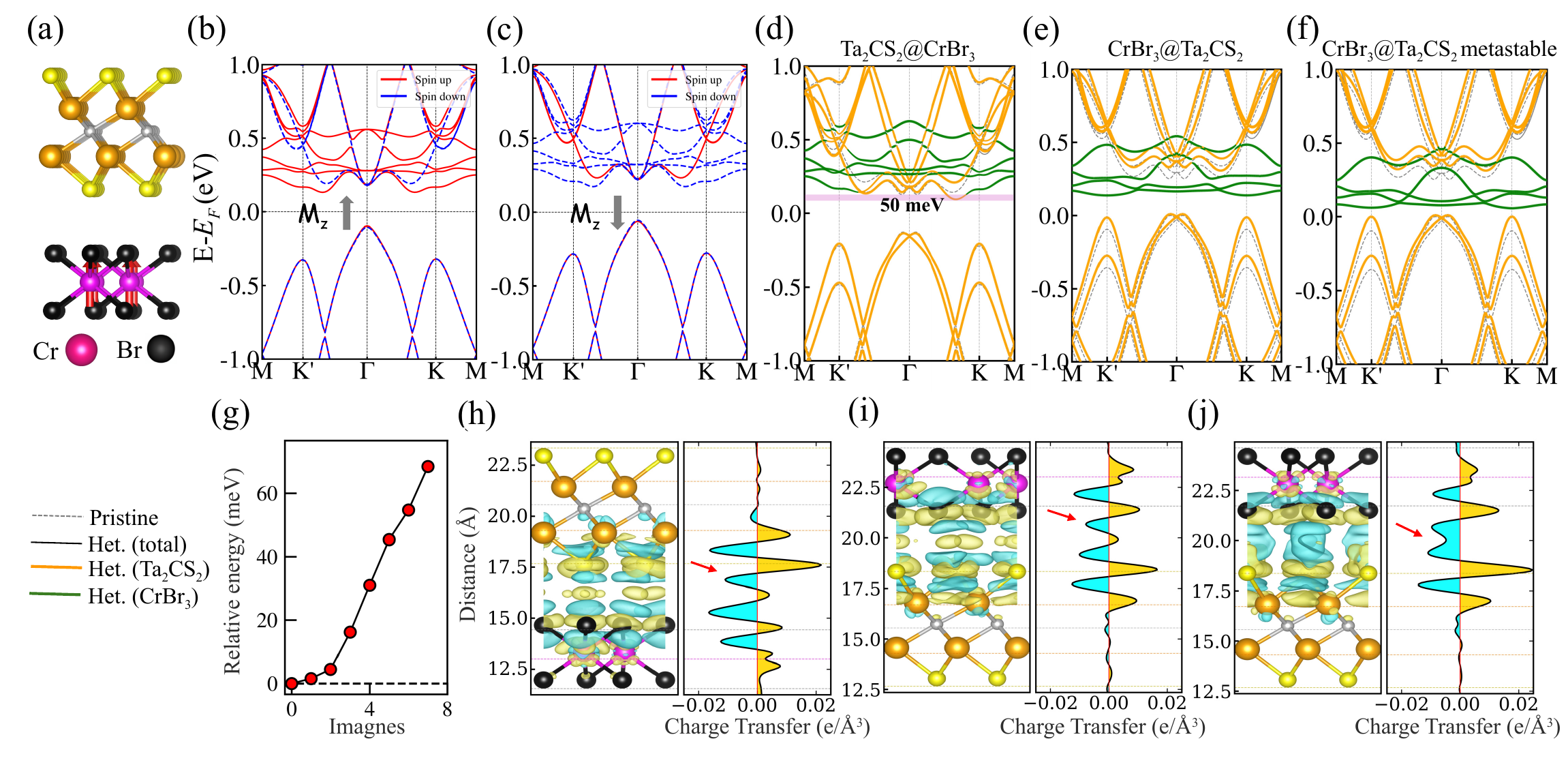}
  \caption{\textbf{Stacking-, magnetization-, and sliding-controlled electronic structure of Ta$_2$CS$_2$ and CrBr$_3$ hetrostructure}.
(a) Side view of Ta$_2$CS$_2$@CrBr$_3$ hetrostructure. (b,c) NSOC band structures for Ta$_2$CS$_2$@CrBr$_3$ heterostructure with magnetic order ($M_z$) along $+z$ and $-z$, respectively. (d) Layer-weighted SOC bands for Ta$_2$CS$_2$@CrBr$_3$, showing a valley splitting of $\sim$50~meV near the conduction-band edge. (e,f) Inverted stacking CrBr$_3$@Ta$_2$CS$_2$: the relaxed (e) and laterally displaced metastable (f) configurations. (g) NEB energy profile connecting the relaxed and metastable top-stacked configurations. (h--j) Charge-density-difference isosurfaces and planar-averaged $\Delta\rho(z)$ profiles for Ta$_2$CS$_2$@CrBr$_3$ (h), CrBr$_3$@Ta$_2$CS$_2$ relaxed (i), and CrBr$_3$@Ta$_2$CS$_2$ metastable (j), highlighting stacking-dependent interfacial charge rearrangement.}
  \label{fig:fig4}
\end{figure*}

Motivated by the pronounced SOC-driven band-edge features in the sulfide MXenes, we examine the impact of a ferromagnetic CrBr$_3$ substrate on their low-energy electronic structure through magnetic proximity. We first establish a commensurate Ta$_2$CS$_2$@CrBr$_3$ heterostructure using a $2\times2$ Ta$_2$CS$_2$ cell matched to a $1\times1$ CrBr$_3$ cell, yielding a lattice mismatch of $\sim$1\%. Three distinct lateral stacking configurations were considered, and the lowest-energy geometry (Fig.~\ref{fig:fig4}a) is used in the analysis below (see Fig.~S6 and Table~S1 for all configurations and their relative energies). Because the ferromagnetic CrBr$_3$ substrate breaks time-reversal symmetry, Ta$_2$CS$_2$ experiences a proximity exchange field whose sign follows the out-of-plane magnetization. Accordingly, reversing $M_z$ reverses the exchange-induced spin polarization: the spin-resolved SOC bands interchange when $M_z$ is flipped (Fig.~\ref{fig:fig4}b,c). This provides a magnetically switchable degree of freedom, enabling deterministic control of spin/valley polarization and suggesting functionality for nonvolatile spin--valley logic, magnetic gating, and proximity-engineered spin filtering in van der Waals heterostructures.

We next analyze the SOC band structure for the Ta$_2$CS$_2$@CrBr$_3$ heterostructure (Fig.~\ref{fig:fig4}d). Layer-weighted bands show that the Ta$_2$CS$_2$-derived valence manifold is only weakly perturbed, while CrBr$_3$-derived states appear near the Ta$_2$CS$_2$ conduction-band edge and renormalize it in a valley-dependent manner: the conduction edge is pushed deeper along $\Gamma$--$K$ than $\Gamma$--$K'$, producing a valley splitting of $\sim$50~meV (Fig.~\ref{fig:fig4}d, inset). The fundamental gap decreases only modestly (from $0.240$ to $0.220$~eV), so the proximity coupling drives valley-selective reshaping rather than a wholesale reconstruction of the Ta$_2$CS$_2$ manifold. This anisotropic CB shift is what enables magnetically-tunable valley-polarized transport in this platform.

\begin{figure*}[t]
  \centering
  % try larger trims; units = left bottom right top
  \includegraphics[
    width=\textwidth,
    height=0.9\textheight,
    keepaspectratio,
    trim=0.8cm 0.2cm 3cm 0.4cm,
    clip
  ]{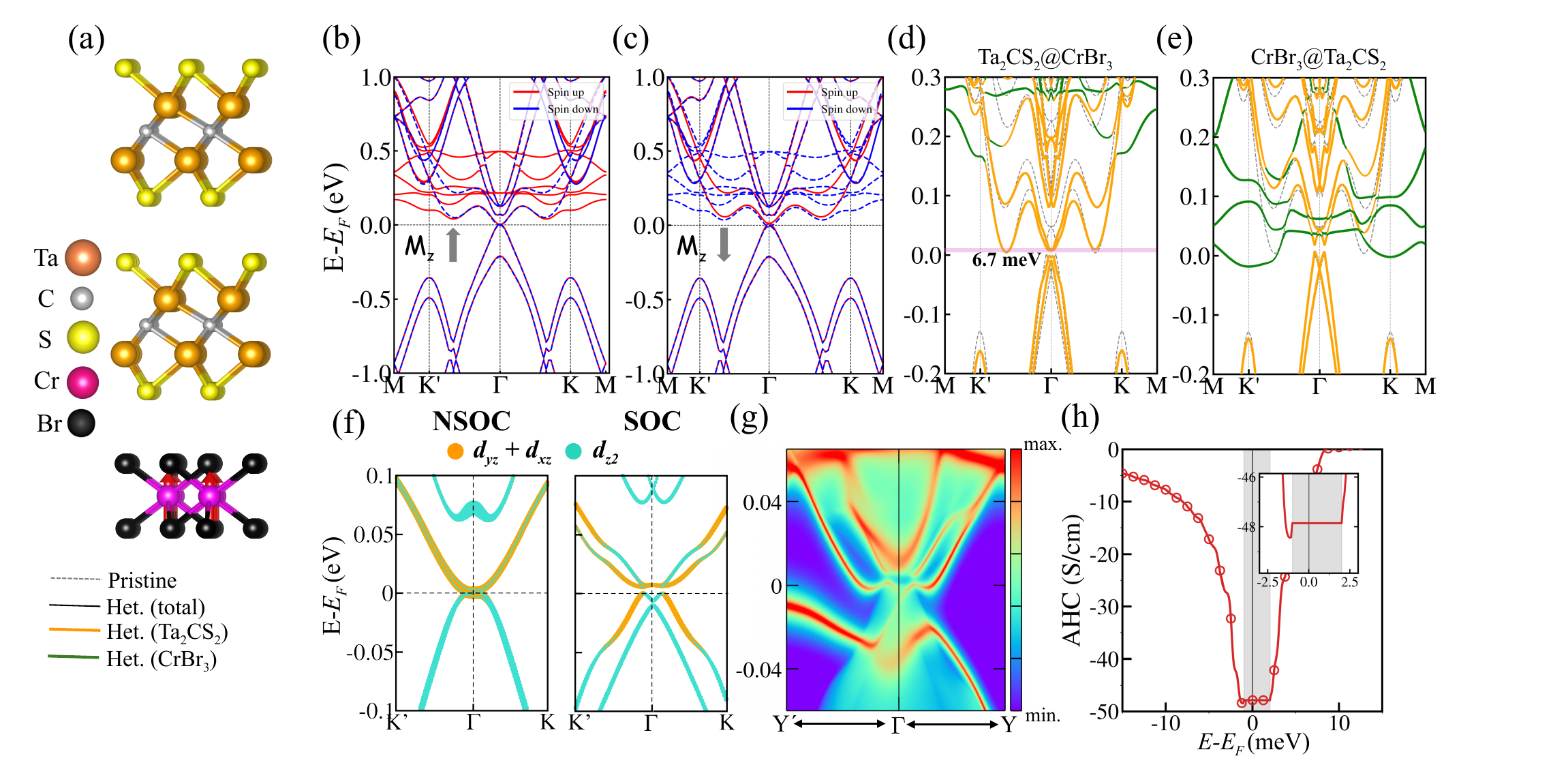}
\caption{\textbf{Proximity-induced magnetism and topological responses in bilayer Ta$_2$CS$_2$@CrBr$_3$}. (a) Side view of the relaxed heterostructure (atom colors as labeled). (b,c) NSOC band structures for out-of-plane CrBr$_3$ magnetization oriented along $+z$ and $-z$. (d) SOC band structure with layer-resolved weights (red: bilayer Ta$_2$CS$_2$; blue: CrBr$_3$), highlighting the Ta$_2$CS$_2$-dominated low-energy bands under magnetic proximity. (e) SOC band structure for inverted stacking with CrBr$_3$ placed above bilayer Ta$_2$CS$_2$. (f) Orbital-resolved bands near the inverted gap without SOC and with SOC, evidencing SOC-driven band inversion within the Ta-$d$ manifold. (g) Edge spectral function showing in-gap boundary modes. (h) Energy-dependent AHC with a magnified view near the Fermi level (inset). (i) OHC $\sigma^{L_z}_{yx}/(2\pi)$ in the vicinity of $E_F$ with the gap scale indicated.}
  \label{fig:fig5}
\end{figure*}

This selectivity has a transparent orbital-symmetry origin. At $\Gamma$, the band-edge states have dominantly $d_{z^2}$ ($L_z=0$) character, on which the substrate exchange field has no orbital coupling. Away from $\Gamma$, both the valence and conduction manifolds along $\Gamma$--K/K$'$ are dominated by $d_{x^2-y^2}\mp i d_{xy}$ ($L_z=\mp 2\hbar$) character (Fig.~S22). The exchange field therefore acts as a $\Delta E \propto B^{\rm eff}_z L_z$ Zeeman-like shift of opposite sign at K and K$'$ on both edges. The asymmetry is nevertheless visible only on the conduction band because the intrinsic atomic-SOC splitting between the $L_z=\pm 2$ branches in the conduction manifold is more than a factor of two smaller than in the deeper valence ($\sim$60~meV vs.\ several hundred meV; Fig.~\ref{fig:fig1}f), so the substrate-induced shift, while present at both edges, is comparable to the intrinsic splitting only at the CB. This is further reinforced by the fact that CrBr$_3$-derived conduction states descend into the Ta$_2$CS$_2$ CB energy window and hybridize with it, whereas the valence manifold lies well below the active CrBr$_3$ states; the hybridization-driven contribution to the valley splitting is therefore present only on the CB side as well.

A complementary, real-space view comes from the atom-resolved spin projections (Fig.~S18). The two Ta sublattices of the polar monolayer host the band manifold unequally (see Fig.~S16), and contact with CrBr$_3$ inverts this assignment: the Ta adjacent to the substrate hybridizes with the Cr-derived conduction states that descend into the relevant energy window, and the resulting bonding--antibonding splitting pushes those Ta states upward into the CB; the far Ta, effectively decoupled, retains its character and acts as the valence-band host. The Rashba winding at $\Gamma$ and the spin--valley locking at K/K$'$ are properties of the orbital character ($d_{z^2}$ and $d_{x^2-y^2}\mp i d_{xy}$, respectively) rather than of the hosting Ta site, so these spin features migrate with the orbitals onto the opposite Ta and are now dressed by the out-of-plane Cr exchange field; consistently, CrBr$_3$ contributes almost exclusively to $\langle S_z \rangle$, acting as a pure Zeeman-like source.

Magnetic proximity also amplifies the nonlinear Hall propensity: for the monolayer heterostructure, $|D|_{\max}$ reaches $\approx 12.9$~\AA{} (Table~\ref{tab:het_summary}, Fig.~S9), more than twice the $5.4$~\AA{} of the pristine monolayer. The enhancement reflects time-reversal-symmetry breaking by the exchange field, which removes the residual K/K$'$ cancellation of Berry-curvature contributions present in the pristine material, together with the valley-selective band-edge shift that relocates Fermi contours into regions of larger $|\partial \Omega_z / \partial k|$.

\begin{table*}[t]
\caption{\textbf{Summary of key electronic and nonlinear-response metrics for the M$_2$CS$_2$@CrBr$_3$ heterostructures}. Reported are the band gap $E_g$, Chern number $C$, valley anisotropy $\Delta_{\mathrm{val}}$, maximum absolute Berry-curvature dipole $|D|_{\max}$, and maximum absolute shift-current response $|\sigma|_{\max}$.}
\label{tab:het_summary}
\centering

\setlength{\tabcolsep}{3pt}
\renewcommand{\arraystretch}{1.10}
\footnotesize

\sisetup{
  table-number-alignment = center,
  table-text-alignment   = center
}

\begin{tabular}{ll
  S[table-format=3.1]
  S[table-format=1.0]
  S[table-format=2.1]
  S[table-format=2.2]
  S[table-format=1.2]
}
\toprule
& & \multicolumn{2}{c}{Gap} & \multicolumn{1}{c}{Valley} & & \\
\cmidrule(lr){3-4}
Material & Thickness
& {$E_g$} & {$C$}
& {$\Delta_{\mathrm{val}}$}
& {$|D|_{\max}$}
& {$|\sigma|_{\max}$} \\
& &
{(meV)} & {}
& {(meV)}
& {(\si{\angstrom})}
& {(\si{\angstrom\milli\ampere\per\volt\squared})} \\
\midrule

Nb$_2$CS$_2$/CrBr$_3$ & Monolayer & 170.5 & 0 & 17.8 & 18.44 & -0.76 \\
Ta$_2$CS$_2$/CrBr$_3$ & Monolayer & 227.5 & 0 & 50.0 & 12.90 & 1.00 \\
Nb$_2$CS$_2$/CrBr$_3$ & Bilayer   & 8.1   & 1 & 0   & {}    & {} \\
Ta$_2$CS$_2$/CrBr$_3$ & Bilayer   & 5.3   & 1 & 6.7  & {}    & {} \\
\bottomrule
\end{tabular}
\end{table*}

We next invert the stacking, placing CrBr$_3$ on top of Ta$_2$CS$_2$ (CrBr$_3$@Ta$_2$CS$_2$) and evaluating three lateral configurations (Fig.~S7, Table~S1). In the lowest-energy top-stacked geometry, the CrBr$_3$-derived conduction manifold shifts downward and the gap reduces to $E_g=0.128$~eV (Fig.~\ref{fig:fig4}e); the metastable configuration drives a further reduction to $E_g=0.047$~eV (Fig.~\ref{fig:fig4}f), approaching a near-semimetallic regime. The Ta$_2$CS$_2$ bands remain comparatively rigid throughout (Fig.~\ref{fig:fig4}e,f), so lateral sliding modulates the gap by tuning the interlayer overlap that sets the CrBr$_3$ conduction edge.

Applying the same sublattice-resolved analysis to the top-stacked geometry (Fig.~S19) shows that the near-Ta/far-Ta rule is universal: irrespective of which face of the Ta$_2$CS$_2$ is in contact with CrBr$_3$, the substrate-near Ta hosts the conduction manifold and the far Ta hosts the valence manifold. The Ta proximity polarization is therefore set by the interface geometry rather than by the absolute orientation of the Cr moments in a particular self-consistent solution: although the bottom- and top-stacked calculations happen to converge to opposite CrBr$_3$ magnetizations (visible as a sign flip of $\langle S_z\rangle$ on the Cr$_{1}$+Cr$_{2}$ row in Figs.~S19 and S18), the Ta spin texture remains pinned to the near/far assignment, mirroring the NSOC band reversal demonstrated in Fig.~\ref{fig:fig4}(b,c). The proximity-induced spin texture is  thus controlled by two independent knobs: a geometric one, which Ta$_2$CS$_2$ face contacts the substrate, and a magnetic one, the direction of the Cr moments. Similar results are obtained for Nb$_2$CS$_2$@CrBr$_3$ (Figs.~S20--S21).

A nudged elastic band (NEB) calculation connecting the stable and metastable top-stacked configurations (Fig.~\ref{fig:fig4}g) gives a maximum barrier of $E^\ddagger=68.4$~meV, indicating that the required lateral displacements are energetically modest and experimentally accessible. The Ta$_2$CS$_2$@CrBr$_3$ gap can therefore be tuned mechanically over a wide range ($0.220$ to $0.047$~eV), providing a route to nanoelectromechanical control of transport in magnetic van der Waals heterostructures.

To distinguish coupling-driven orbital reorganization from a trivial macroscopic electrostatic shift, we analyzed the planar-averaged charge-density difference (CDD), $\Delta\rho(z) = \rho_{\mathrm{het}}(z) - \rho_{\mathrm{Ta_2CS_2}}(z) - \rho_{\mathrm{CrBr_3}}(z)$, for the three stacking configurations (Fig.~\ref{fig:fig4}h--j). All three exhibit a clear interfacial polarization, with electron accumulation on either side of the vdW gap and depletion within it. Upon stacking inversion and subsequent lateral translation, the depletion region becomes increasingly localized while the flanking accumulation peaks intensify; in the metastable configuration, the two depletion lobes merge and the 
central accumulation feature seen in the bottom-stacked case is strongly suppressed. This evolution is characteristic of enhanced short-range overlap and a Pauli-repulsion (``pushback'') response, signaling progressively stronger configuration-dependent interlayer coupling.

To rule out a macroscopic electrostatic origin of the gap modulation, we extracted the vacuum-referenced work function $\Phi = V_{\mathrm{vac}} - E_F$ for each configuration 
(Table~S3). Two observations are decisive. Stacking inversion (bottom $\rightarrow$ top) increases $\Phi$ by $\sim 64$~meV, consistent with a reorientation of the interface dipole when the vacuum-facing termination is exchanged. The two top-stacked configurations, however, differ in $\Phi$ by only $\sim 8$~meV, while their gaps differ by $\Delta E_g = 81.2$~meV, an order of magnitude larger. The sliding-induced gap reduction is therefore not an electrostatic effect; it arises from configuration-dependent interlayer hybridization that selectively renormalizes the CrBr$_3$ conduction edge while the Ta$_2$CS$_2$ manifold remains rigid.

\begin{figure*}[t]
  \centering
  % try larger trims; units = left bottom right top
  \includegraphics[
    width=\textwidth,
    height=0.9\textheight,
    keepaspectratio,
    trim=0.1cm 0cm 0.1cm 0cm,
    clip
  ]{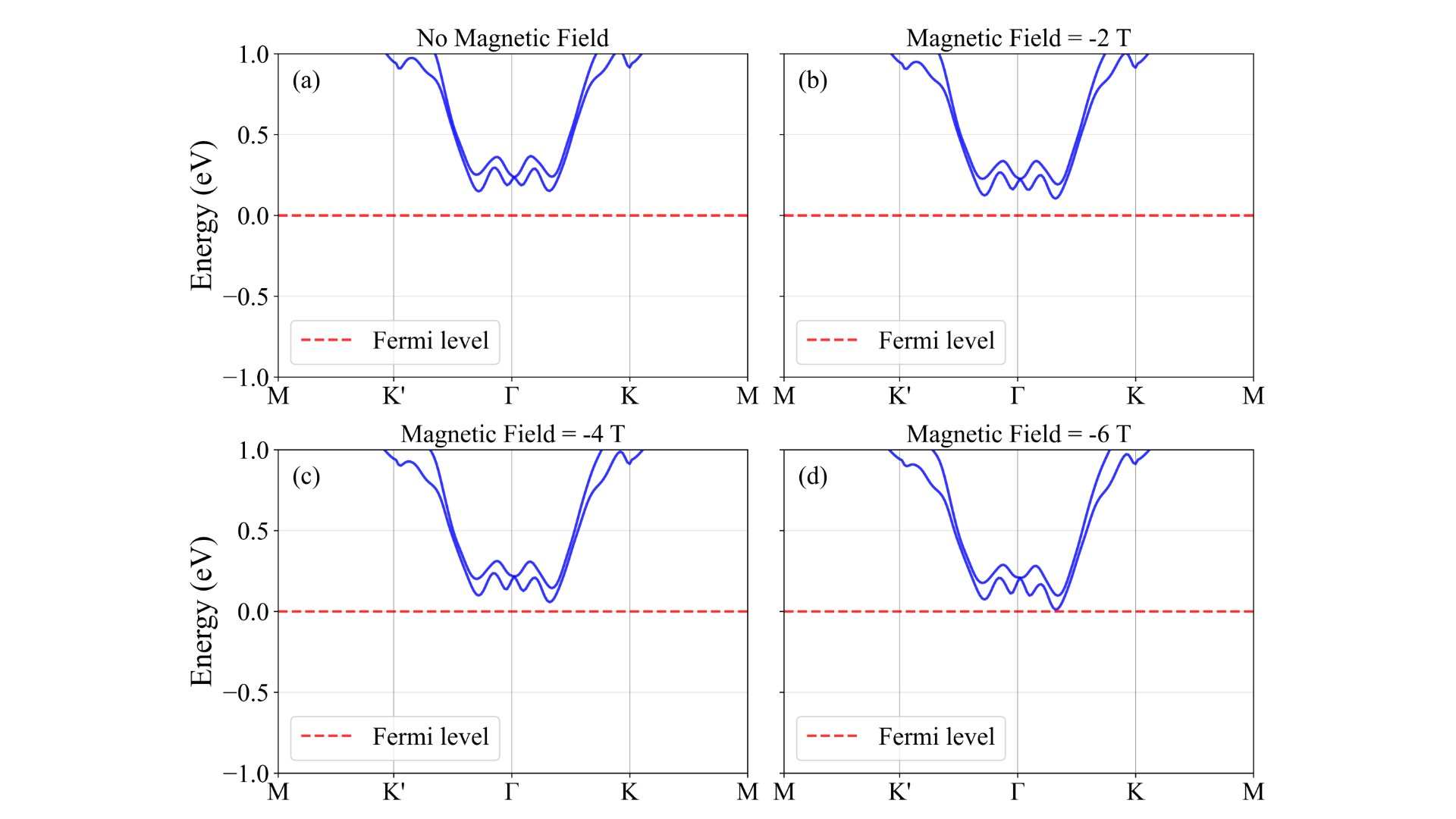}
\caption{\textbf{Band structures of the 4-band tight-binding model for Ta$_2$CS$_2$ monolayer under varying magnetic fields applied to conduction bands:} (a) No magnetic field, (b) Magnetic field = $-2$ T, (c) Magnetic field = $-4$ T, (d) Magnetic field = $-6$ T. The Fermi level is indicated by the dashed red line.}
  \label{fig:fig6}
\end{figure*}

We next consider the bilayer Ta$_2$CS$_2$@CrBr$_3$ heterostructure (Fig.~\ref{fig:fig5}a; tested stacking configurations and relative energies in Fig.~S10 and Table~S2). As in the monolayer case, reversing the out-of-plane magnetization of CrBr$_3$ flips the sign of the proximity exchange field, and the spin-resolved band features interchange between the two M$_z$ orientations (Fig.~\ref{fig:fig5}b,c). The low-energy bands around $\Gamma$ remain dominated by the bilayer Ta$_2$CS$_2$ manifold, with a small SOC-induced gap of $E_g = 5.3$~meV (Table~\ref{tab:het_summary}).

Layer-weighted SOC bands (Fig.~\ref{fig:fig5}d) show that the bilayer Ta$_2$CS$_2$ states dominate both band edges, with the CrBr$_3$-derived conduction manifold pushed energetically further from $\Gamma$ than in the monolayer heterostructure. The proximity influence is correspondingly reduced: the CB shifts slightly more along K$'$--$\Gamma$ than along $\Gamma$--K (opposite to the monolayer case), yielding a valley splitting of only $\sim$6.7~meV. Despite this weak proximity coupling, the SOC-induced gap originates from the same Ta-$d$ band inversion between $d_{xz}+d_{yz}$ and $d_{z^2}$ manifolds that makes the pristine bilayer a quantum spin Hall insulator (Fig.~\ref{fig:fig5}f, cf.\ Fig.~\ref{fig:fig3}b); time-reversal symmetry breaking by CrBr$_3$ converts this $Z_2$ phase into its quantum anomalous Hall descendant. The calculated Chern number is $C=1$: consistently, the ribbon spectrum hosts a single chiral gap-spanning edge mode (Fig.~\ref{fig:fig5}g), and the anomalous Hall conductivity exhibits a quantized plateau within the gap (Fig.~\ref{fig:fig5}h).

In the inverted stacking, CrBr$_3$@Ta$_2$CS$_2$ (Fig.~\ref{fig:fig5}e), the CrBr$_3$ conduction manifold shifts downward and hybridizes strongly with Ta$_2$CS$_2$ states near 
the Fermi level, closing the gap and producing a metallic band alignment. The QAH phase of the bottom-stacked geometry is therefore destroyed by stacking inversion, providing another illustration of how strongly the heterostructure electronic structure responds to interfacial geometry.

Nb$_2$CS$_2$@CrBr$_3$ follows the same qualitative trends as the Ta-based stack but with reduced valley anisotropy (Fig.~S8, Table~\ref{tab:het_summary}): the monolayer heterostructure exhibits a CB valley splitting of $\Delta_{\mathrm{val}}=17.8$~meV, an amplified Berry-curvature dipole of $|D|_{\max}=18.44$~\AA, and a shift-current extremum of $|\sigma|_{\max}=0.76$~\AA\,mA/V$^{2}$. The bilayer Nb$_2$CS$_2$@CrBr$_3$ collapses into an ultranarrow-gap regime ($E_g=8.1$~meV; Fig.~S12) with $C=1$, showing that the magnetism-induced QAH phase is not specific to the Ta-based bilayer.

To rationalize the valley-selective conduction-band (CB) reshaping induced by CrBr$_3$, we constructed a minimal Wannier tight-binding (TB) model from maximally localized Wannier functions and added an effective proximity/exchange term that acts predominantly within the CB-like subspace. Specifically, starting from the Wannier Hamiltonian
\begin{equation}
H_0(\mathbf{k})=\sum_{\mathbf{R}} e^{i\mathbf{k}\cdot\mathbf{R}}\,H(\mathbf{R}),
\end{equation}
we introduce a phenomenological, magnetization-induced shift applied to the CB-like Wannier orbitals,
\begin{equation}
H(\mathbf{k}) = H_0(\mathbf{k}) + \Delta_{\mathrm{prox}}(\mathbf{k}) \sum_{n\in \mathrm{CB}} |n\rangle\langle n|,
\end{equation}
where the projector restricts the perturbation to the chosen CB-like Wannier orbitals to describe the evolution of the conduction bands away from the $\Gamma$ point. Figure~\ref{fig:fig6} compares the resulting CB dispersions for different values of an effective proximity parameter $B_{\mathrm{eff}}$ (used here as a convenient tuning knob). As $|B_{\mathrm{eff}}|$ increases, the CB edge exhibits a systematic downward renormalization that is anisotropic between the two $\Gamma$--K lines: the shift along $\Gamma$--K is enhanced relative to $\Gamma$--K$^\prime$, generating an energy difference between the two directions comparable to the $\sim$50~meV splitting obtained from the DFT heterostructure. 
Note that the present 4-band model is a low-energy effective description focused on the conduction bands near $\Gamma$; it does not explicitly include the $K$/$K'$ points and a larger orbital basis would be required to quantitatively capture the full valley physics. Overall, this minimal Wannier-based framework demonstrates that a momentum-dependent, CB-targeted proximity perturbation is sufficient to reproduce the qualitative trend observed in Ta$_2$CS$_2$@CrBr$_3$, yielding sizable momentum-dependent anisotropy without invoking a wholesale reconstruction of the Ta$_2$CS$_2$ valence bands.

In summary, we have established chalcogen-terminated vdW MXenes (M$_2$CX$_2$; M=Ta, Nb; X=S, Se) as a versatile and multifunctional platform for engineering spin–valley and topological phenomena. Our results demonstrate that these materials host some of the most pronounced SOC-driven effects reported to date in natural 2D systems, characterized by a giant Rashba parameter of $2.53 \text{ eV\AA}$ and a momentum-selective concentration of Berry curvature. These features drive exceptional second-order responses, with an intrinsic shift current reaching $|\sigma|_{\max} \approx 5~\text{\AA\,mA/V}^2$ and a Berry-curvature dipole attaining $|D|_{\max} \approx 18.44$~\AA\ in the heterostructure limit, positioning these MXenes as premier candidates for junction-free photovoltaics and nonlinear Hall electronics. Beyond the monolayer limit, we identify a topological transition in the pristine M$_2$CS$_2$ bilayers, where an SOC-gapped M-$d$ band inversion produces a nontrivial $Z_2=1$ quantum spin Hall phase. By interfacing these MXenes with ferromagnetic $\text{CrBr}_3$, we achieve deterministic magnetic control over the spin and valley degrees of freedom through a magnetization-reversible proximity exchange field. Furthermore, we demonstrate that the electronic bandgap is highly susceptible to mechanical tuning; both stacking inversion and lateral sliding serve as efficient levers to modulate the exchange-SOC interplay, which additionally yields an emergent $C=1$ quantum anomalous Hall phase in the bilayer heterostructure. Collectively, our findings provide a comprehensive roadmap for the experimental realization of vdW MXene/magnet stacks. By integrating magnetic, topological, and nanoelectromechanical control, these heterostructures offer a transformative route toward switchable valleytronics, high-performance nonlinear optoelectronics, and dissipationless quantum devices in the atomically thin limit.

\section*{Acknowledgements}

This work was partially supported by the Wallenberg Initiative Materials Science for Sustainability (WISE)
funded by the Knut and Alice Wallenberg Foundation.
EvL acknowledges support from the Swedish Research Council (Vetenskapsrådet, VR) under grant 2022-03090, from the Royal Physiographic Society in Lund and by eSSENCE, a strategic research area for e-Science, grant number eSSENCE@LU 9:1. J.A.L. thank the Kempe-stiftelserna, Sweden and Swedish Research Council under grant no. 2023-03894 for financial support.
The computations were enabled by resources provided by the National Academic Infrastructure for Supercomputing in Sweden (NAISS), partially funded by the Swedish Research Council through grant agreement no. 2022-06725.

%%\printbibliography

%%\bibliography{references.bib}

\input{main.bbl}

% ============================================================
% SUPPLEMENTAL MATERIAL
% ============================================================
 
\clearpage
\onecolumn
 
\setcounter{page}{1}
\setcounter{table}{0}
\setcounter{figure}{0}
\setcounter{equation}{0}
\setcounter{section}{0}
 
\renewcommand{\thepage}{S\arabic{page}}
\renewcommand{\thetable}{S\arabic{table}}
\renewcommand{\thefigure}{S\arabic{figure}}
\renewcommand{\theequation}{S\arabic{equation}}
\renewcommand{\thesection}{S\arabic{section}}
 
\begin{center}
    {\Large\textbf{Supplemental Material for}} \\[0.8em]
    {\large\itshape Giant Rashba Splitting and Enhanced Nonlinear Berry-Phase Responses in Sliding-Tunable vdW MXene Heterostructures} \\[1.8em]
 
    Ali Sufyan,$^{1,2,3}$ J. Andreas Larsson,$^{3,4}$ Andreas Kreisel,$^{5}$ and Erik van Loon$^{1,2,6}$* \\[1.2em]
 
    {\small
    $^{1}$NanoLund and Division of Mathematical Physics, Department of Physics, Lund University, SE-221 00 Lund, Sweden \\
    $^{2}$Wallenberg Initiative Materials Science for Sustainability, Department of Physics, Lund University, SE-221 00 Lund, Sweden \\
    $^{3}$Applied Physics, Division of Materials Science, Department of Engineering Sciences and Mathematics, Lule\aa\ University of Technology, SE-971 87 Lule\aa, Sweden \\
    $^{4}$Wallenberg Initiative Materials Science for Sustainability, Lule\aa\ University of Technology, SE-971 87 Lule\aa, Sweden \\
    $^{5}$Department of Physics and Astronomy, Box 524, SE-751 20 Uppsala, Sweden \\
    $^{6}$LINXS Institute of advanced Neutron and X-ray Science (LINXS), Lund, Sweden
    }
\end{center}
 
\vspace{2em}
\noindent\rule{\textwidth}{0.4pt}
\vspace{1em}

\section{Computational Details}
All first-principles calculations were performed within density functional theory (DFT) using the Vienna \emph{ab initio} Simulation Package (VASP)\cite{paw1,paw2,vasp}. The exchange--correlation functional was treated within the generalized gradient approximation (GGA) in the Perdew--Burke--Ernzerhof (PBE) form\cite{perdew1996generalized}. On-site Coulomb interactions were included using the GGA+$U$ approach with an effective Hubbard parameter $U_{\mathrm{eff}}=3$~eV applied to the Ta-$d$ and Cr-$d$ states\cite{zhang2020abundant,wu2024chirality,moore2024high}. Spin--orbit coupling (SOC) was included self-consistently for all electronic-structure calculations reported in the main text unless stated otherwise. The plane-wave kinetic-energy cutoff was set to 550~eV. Long-range dispersion interactions between layers were accounted for using the DFT-D3 method\cite{grimme2006semiempirical}. A vacuum region of 20~\AA\ was introduced along the out-of-plane direction to avoid spurious interactions between periodic images.

Monolayer and bilayer M$_2$CX$_2$ structures were modeled using periodic slab geometries and sampled with a $\Gamma$-centered $21\times21\times1$ \textbf{k}-mesh. For the M$_2$CX$_2$/CrBr$_3$ heterostructures, a $\Gamma$-centered $12\times12\times1$ \textbf{k}-mesh was used. Structural relaxations were performed until the residual forces were below $10^{-2}$~eV/{\AA}. For selected monolayer and bilayer cases, we further validated the electronic structure using hybrid-functional calculations within the HSE06 formalism\cite{heyd2003hybrid}, confirming the robustness of the SOC-induced band splittings and near-gap band ordering.
Spin-projected band structures and momentum-resolved spin textures were analyzed using \textsc{pyprocar} based on VASP outputs\cite{herath2020pyprocar,lang2024expanding}. Maximally localized Wannier functions (MLWFs) were constructed from the DFT wavefunctions using \textsc{Wannier90}\cite{marzari1997maximally, pizzi2020wannier90}, yielding tight-binding Hamiltonians employed for topological and Berry-phase analyses. The intrinsic anomalous Hall conductivity (AHC), orbital Hall conductivity (OHC), edge spectra, and topological invariants (including the $Z_2$ index and Chern number) were computed from the Wannier tight-binding models using \textsc{WannierTools}\cite{wu2018wanniertools}. Nonlinear Berry-phase responses, including the Berry-curvature dipole (BCD) and shift-current tensor, were calculated using \textsc{WannierBerri}\cite{tsirkin2021high}. For the BCD and shift-current calculations, we employed a dense Brillouin-zone sampling of $300\times300\times1$ to ensure numerical convergence of the response tensors. Additional structural and electronic analysis and workflow utilities were performed with \textsc{VASPKIT}.

\begin{figure*}[t]
  \centering
  % try larger trims; units = left bottom right top
  \includegraphics[
    width=\textwidth,
    height=0.9\textheight,
    keepaspectratio,
    trim=0.0cm 0.0cm 0.0cm 0.0cm,
    clip
  ]{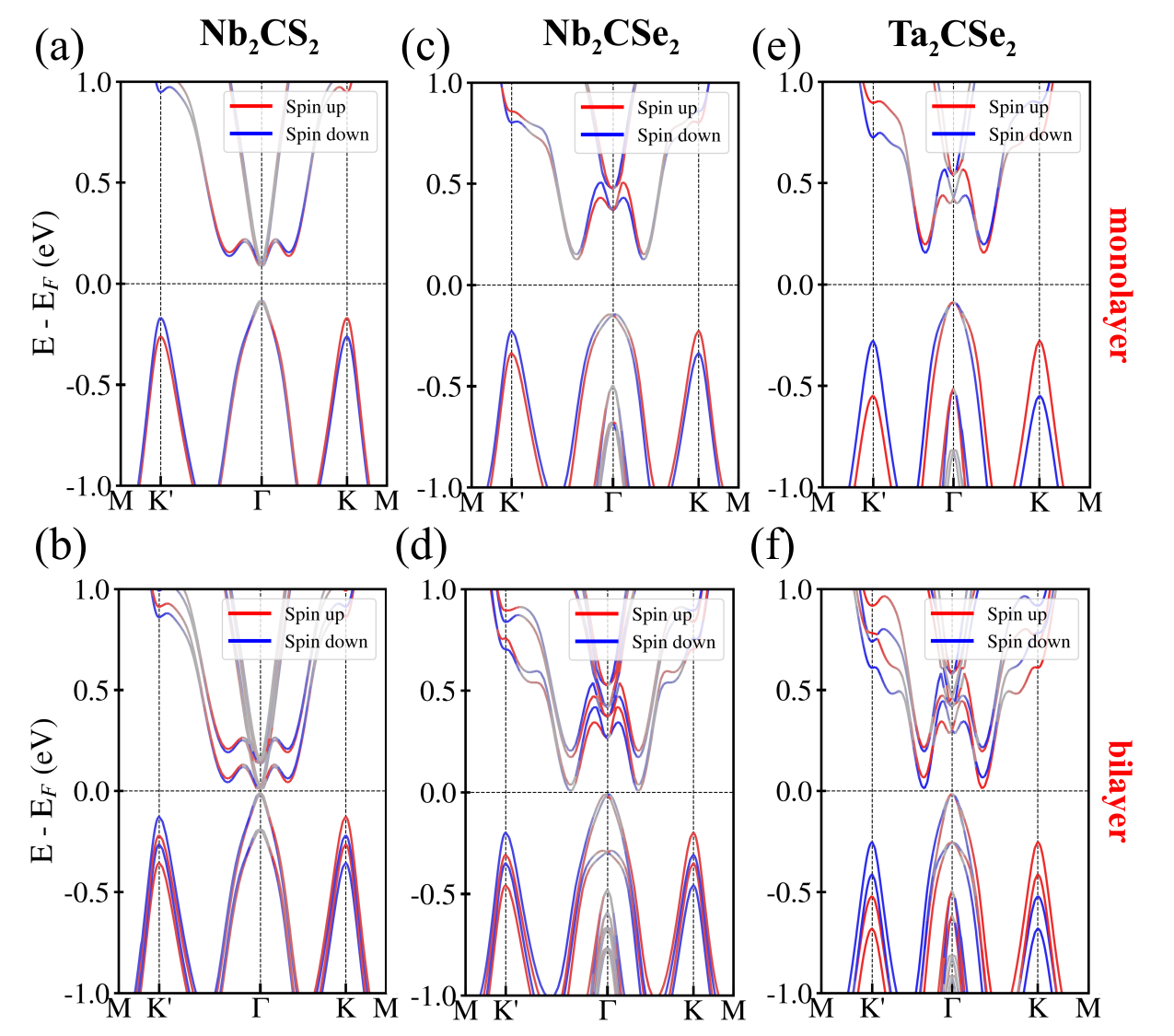}
  \caption{\textbf{SOC band structures of vdW MXenes.}
Spin--orbit-coupled band structures of Nb$_2$CS$_2$ (a,b), Nb$_2$CSe$_2$ (c,d), and Ta$_2$CSe$_2$ (e,f) for monolayers (top row) and bilayers (bottom row), colored by the out-of-plane spin projection $\langle S_z \rangle$ (color scale $\pm 1/2$). }
   \label{fig:S1}
 \end{figure*}

\begin{figure*}[t]
  \centering
  % try larger trims; units = left bottom right top
  \includegraphics[
    width=\textwidth,
    height=0.9\textheight,
    keepaspectratio,
    trim=0.3cm 1.0cm 2.5cm 0.5cm,
    clip
  ]{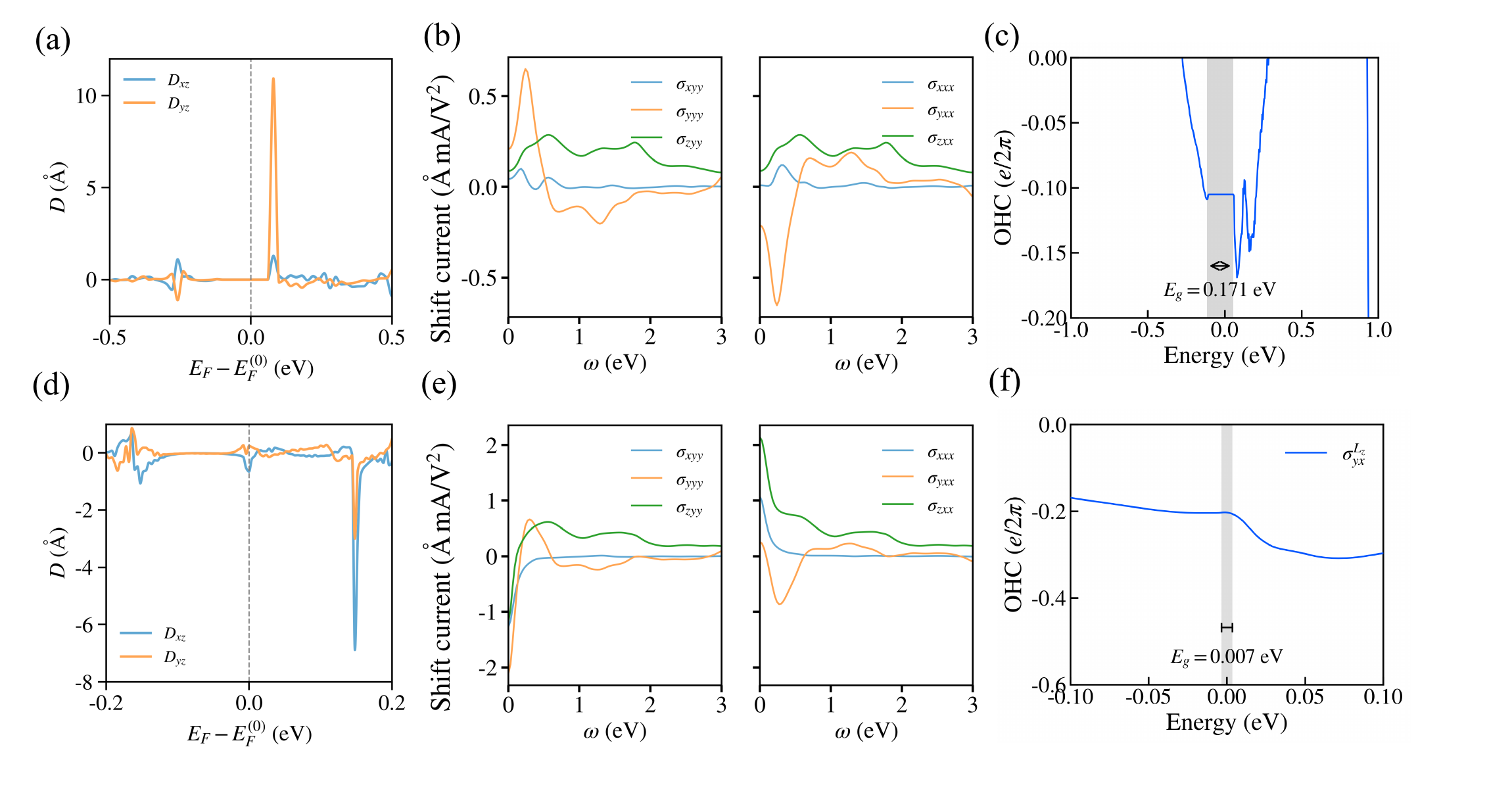}
  \caption{\textbf{Berry-phase responses of Nb$_2$CS$_2$ with SOC.}
(a--c) Monolayer and (d--f) bilayer results: (a,d) Berry-curvature-dipole components $D_a$,
(b,e) frequency-dependent shift-current tensor components $\sigma_{abc}(\omega)$, and
(c,f) energy-dependent orbital Hall conductivity $\sigma^{L_z}_{yx}$, with the bulk band gap $E_g$ indicated.}
   \label{fig:S2}
 \end{figure*}

 \begin{figure*}[t]
  \centering
  % try larger trims; units = left bottom right top
  \includegraphics[
    width=\textwidth,
    height=0.9\textheight,
    keepaspectratio,
    trim=0.3cm 1.0cm 2.5cm 0.5cm,
    clip
  ]{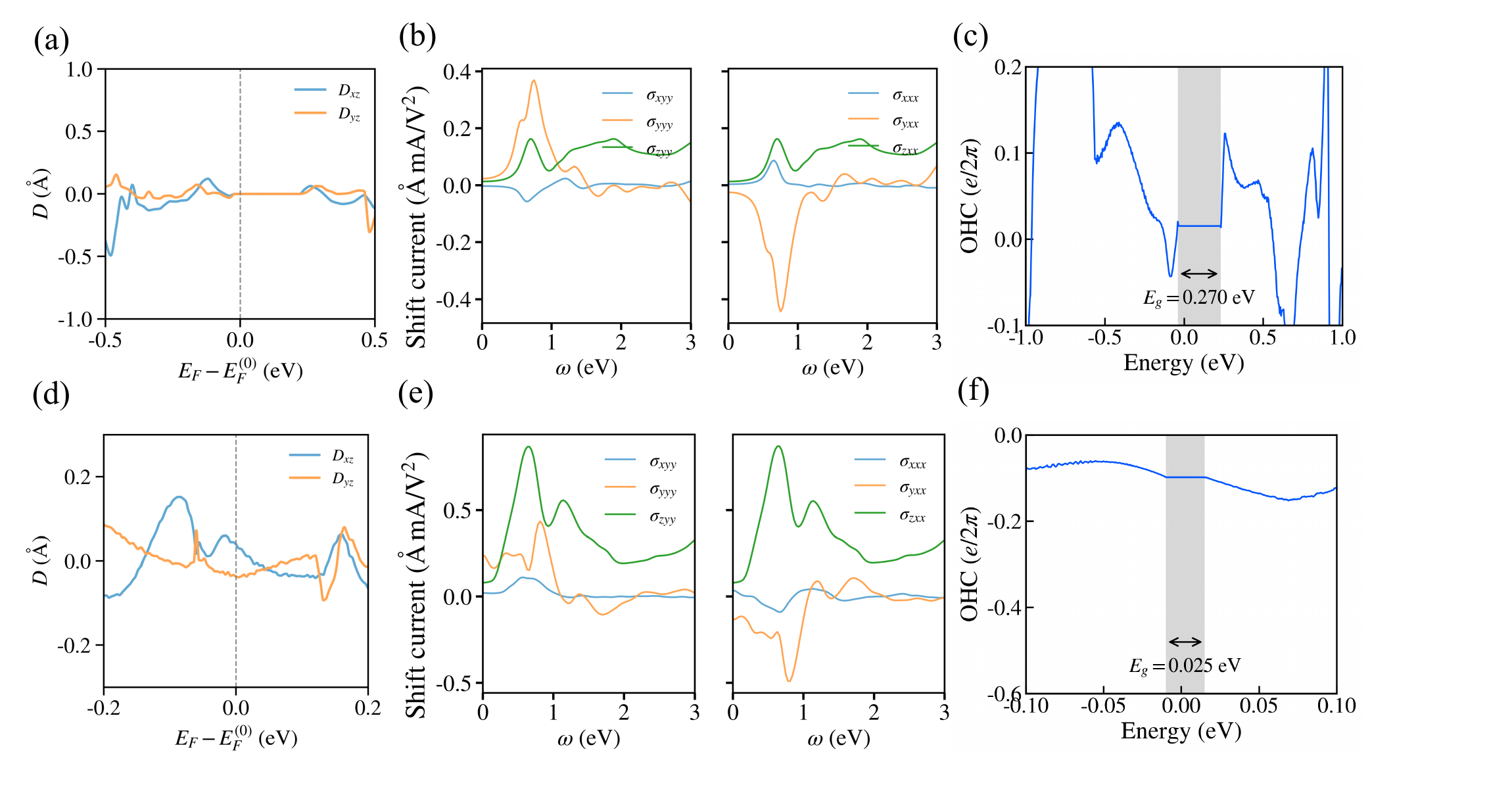}
  \caption{\textbf{Berry-phase responses of Nb$_2$CSe$_2$ with SOC.}
(a--c) Monolayer and (d--f) bilayer results: (a,d) Berry-curvature-dipole components $D_a$,
(b,e) frequency-dependent shift-current tensor components $\sigma_{abc}(\omega)$, and
(c,f) energy-dependent orbital Hall conductivity $\sigma^{L_z}_{yx}$, with the bulk band gap $E_g$ indicated.}
   \label{fig:S3}
 \end{figure*}

\begin{figure*}[t]
  \centering
  % try larger trims; units = left bottom right top
  \includegraphics[
    width=\textwidth,
    height=0.9\textheight,
    keepaspectratio,
    trim=0.3cm 1.0cm 2.5cm 0.5cm,
    clip
  ]{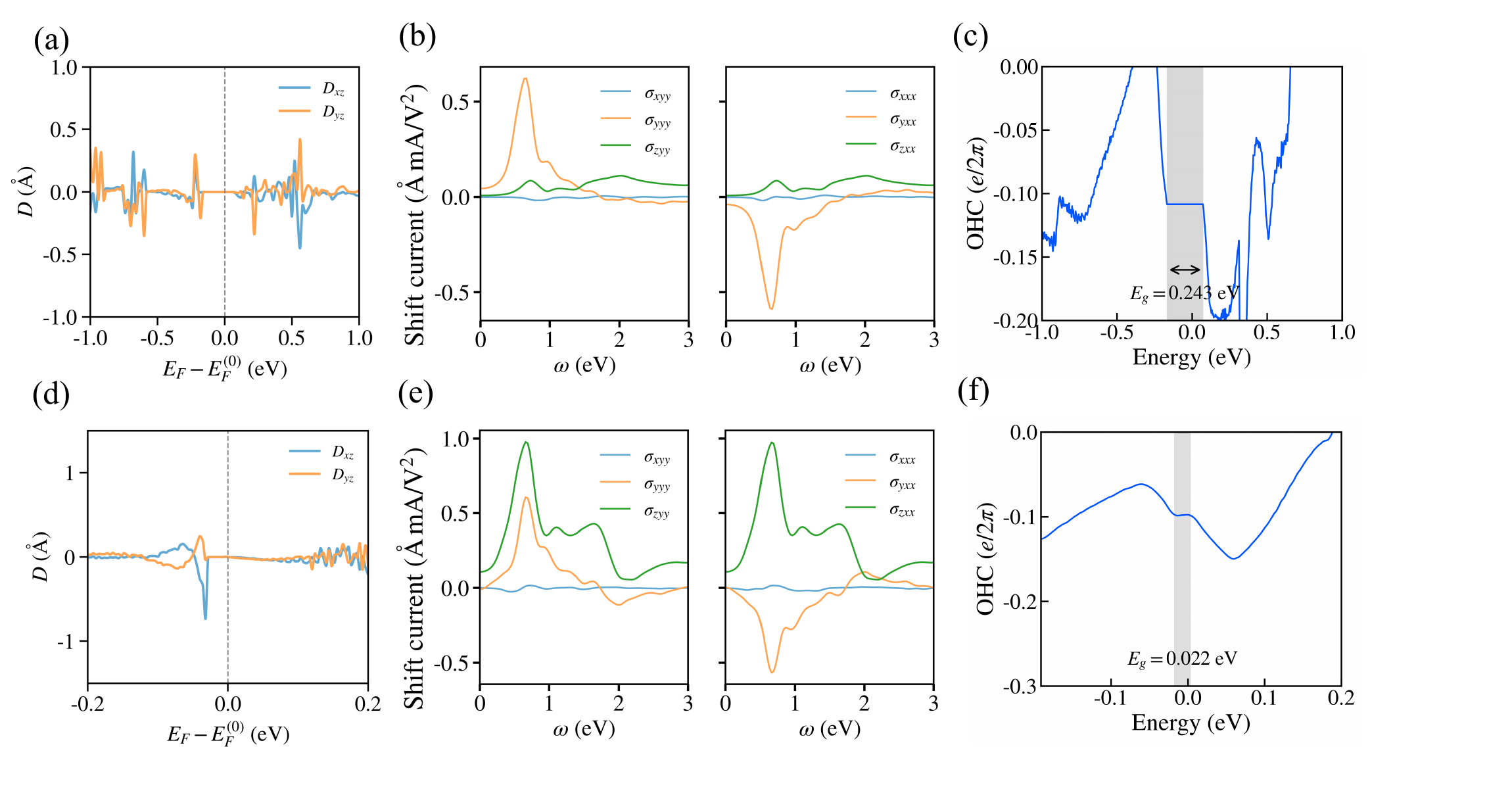}
  \caption{\textbf{Berry-phase responses of Ta$_2$CSe$_2$ with SOC.}
(a--c) Monolayer and (d--f) bilayer results: (a,d) Berry-curvature-dipole components $D_a$,
(b,e) frequency-dependent shift-current tensor components $\sigma_{abc}(\omega)$, and
(c,f) energy-dependent orbital Hall conductivity $\sigma^{L_z}_{yx}$, with the bulk band gap $E_g$ indicated.}
   \label{fig:S4}
 \end{figure*}

 \begin{figure*}[t]
  \centering
  % try larger trims; units = left bottom right top
  \includegraphics[
    width=\textwidth,
    height=0.9\textheight,
    keepaspectratio,
    trim=0.3cm 0.5cm 1cm 0.5cm,
    clip
  ]{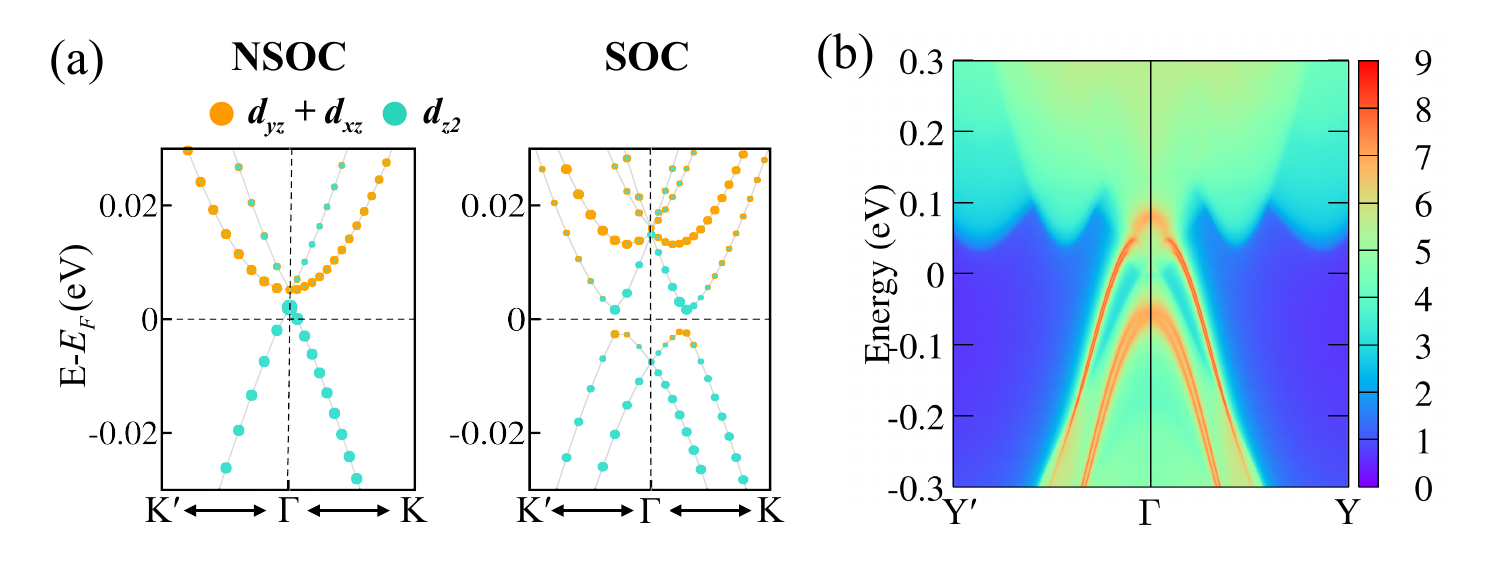}
 \caption{\textbf{Quantum spin Hall state in bilayer Nb$_2$CS$_2$}. (a) Orbital-resolved bands near $\Gamma$ without SOC and with SOC, highlighting the SOC-induced band inversion between Nb-$d$ manifolds. (b) Edge spectral function/band structure of a semi-infinite ribbon, showing gapless helical edge modes connecting valence and conduction continua.}
   \label{fig:S5}
 \end{figure*}

\begin{table}[H]
\caption{Relative energies (meV) of different stacking configurations of monolayer M$_2$CS$_2$/CrBr$_3$ heterostructures, referenced to the most stable configuration in each case.}
\label{tab:relE_stacking}
\centering
\begin{tabular}{l c c c}    %%  <-- changed from "l c S S" to "l c c c"
\toprule
\multirow{2}{*}{System} & \multirow{2}{*}{Config.} & \multicolumn{2}{c}{$\Delta E$ (meV)} \\
\cmidrule(lr){3-4}
 & & CrBr$_3$ bottom & CrBr$_3$ top \\   %% removed the {} braces too
\midrule
\multirow{3}{*}{Ta$_2$CS$_2$}
& I   & 0     & 80.51 \\
& II  & 8.64  & 14.64 \\
& III & 26.17 & 0     \\
\midrule
\multirow{3}{*}{Nb$_2$CS$_2$}
& I   & 0     & 62.72 \\
& II  & 11.32 & 0     \\
& III & 28.23 & 1.56  \\
\bottomrule
\end{tabular}
\end{table}

\begin{figure*}[t]
  \centering
  % try larger trims; units = left bottom right top
  \includegraphics[
    width=\textwidth,
    height=0.9\textheight,
    keepaspectratio,
    trim=0.0cm 5.0cm 0.0cm 1.5cm,
    clip
  ]{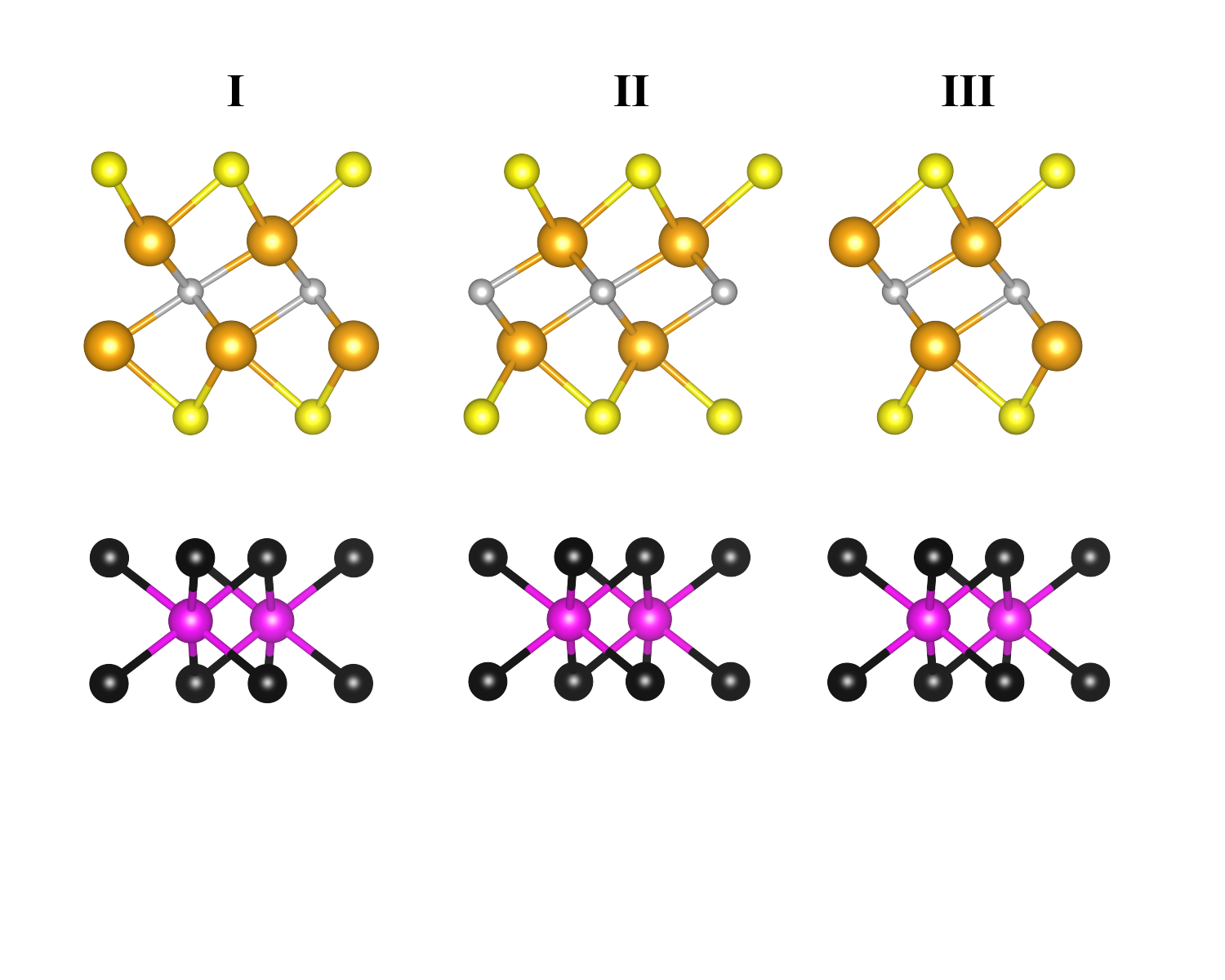}
 \caption{\textbf{Stacking configurations of monolayer M$_2$CS$_2$/CrBr$_3$ heterostructure.}
Side views of the three representative lateral configrations (I--III) considered in this work. Orange/gray/yellow spheres denote M/C/S atoms in M$_2$CS$_2$, and magenta/black spheres denote Cr/Br atoms in CrBr$_3$.}
   \label{fig:S6}
 \end{figure*}

\begin{figure*}[t]
  \centering
  % try larger trims; units = left bottom right top
  \includegraphics[
    width=\textwidth,
    height=0.9\textheight,
    keepaspectratio,
    trim=0.0cm 3.5cm 0.0cm 0.5cm,
    clip
  ]{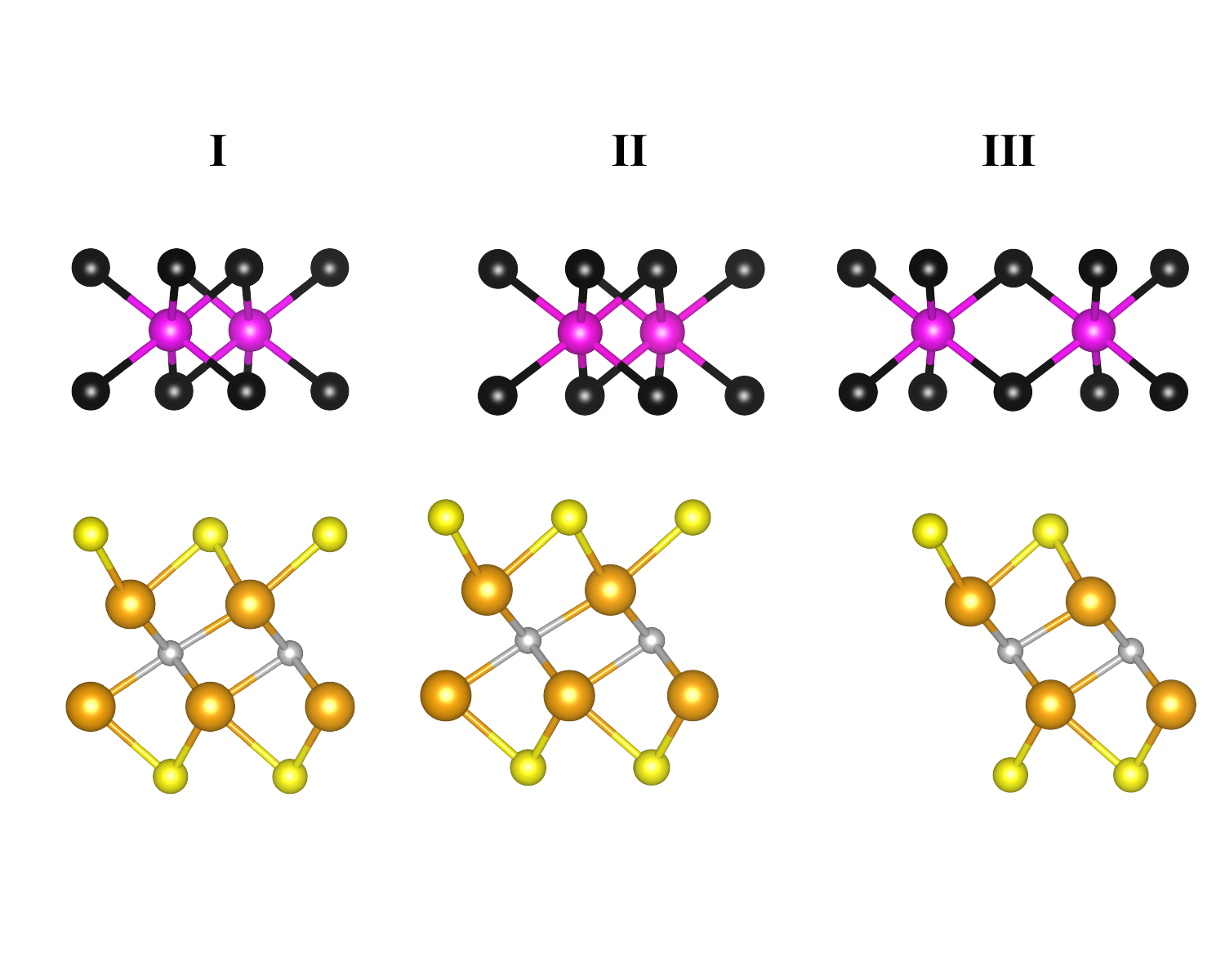}
 \caption{\textbf{Stacking configurations of monolayer CrBr$_3$/M$_2$CS$_2$ heterostructure.}
Side views of the three representative lateral configrations (I--III) considered in this work. Orange/gray/yellow spheres denote M/C/S atoms in M$_2$CS$_2$, and magenta/black spheres denote Cr/Br atoms in CrBr$_3$.}
   \label{fig:S7}
 \end{figure*}

\begin{figure*}[t]
  \centering
  % try larger trims; units = left bottom right top
  \includegraphics[
    width=\textwidth,
    height=0.9\textheight,
    keepaspectratio,
    trim=1.2cm 0.5cm 4.0cm 0.0cm,
    clip
  ]{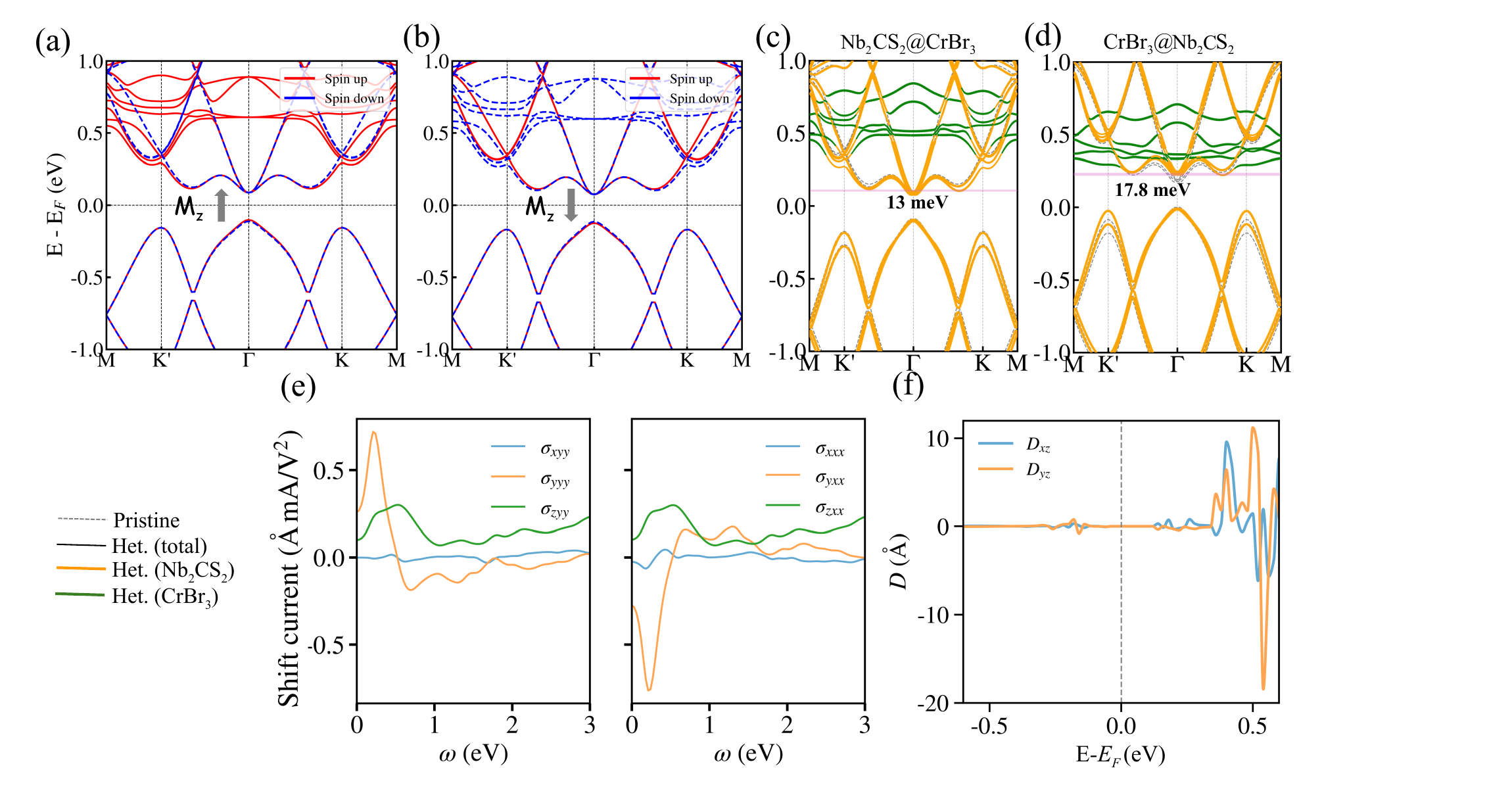}
\caption{\textbf{Stacking-, magnetization-, and sliding-controlled electronic structure of Nb$_2$CS$_2$ and CrBr$_3$ hetrostructure}.
 (a,b) NSOC band structures for Nb$_2$CS$_2$@CrBr$_3$ heterostructure with magnetic order ($M_z$) along $+z$ and $-z$, respectively. (c) Layer-weighted SOC bands for Nb$_2$CS$_2$@CrBr$_3$, showing a valley splitting of $\sim$13~meV near the conduction-band edge. (d) Inverted stacking CrBr$_3$@Nb$_2$CS$_2$: the relaxed configuration, showing a valley splitting of $\sim$17~meV near the conduction-band edge. (e) Frequency-dependent shift-current tensor components $\sigma_{abc}(\omega)$ and (f) BCD coefficient. }
   \label{fig:S8}
 \end{figure*}

\begin{figure*}[t]
  \centering
  % try larger trims; units = left bottom right top
  \includegraphics[
    width=\textwidth,
    height=0.9\textheight,
    keepaspectratio,
    trim=0.0cm 1.0cm 0.0cm 0.0cm,
    clip
  ]{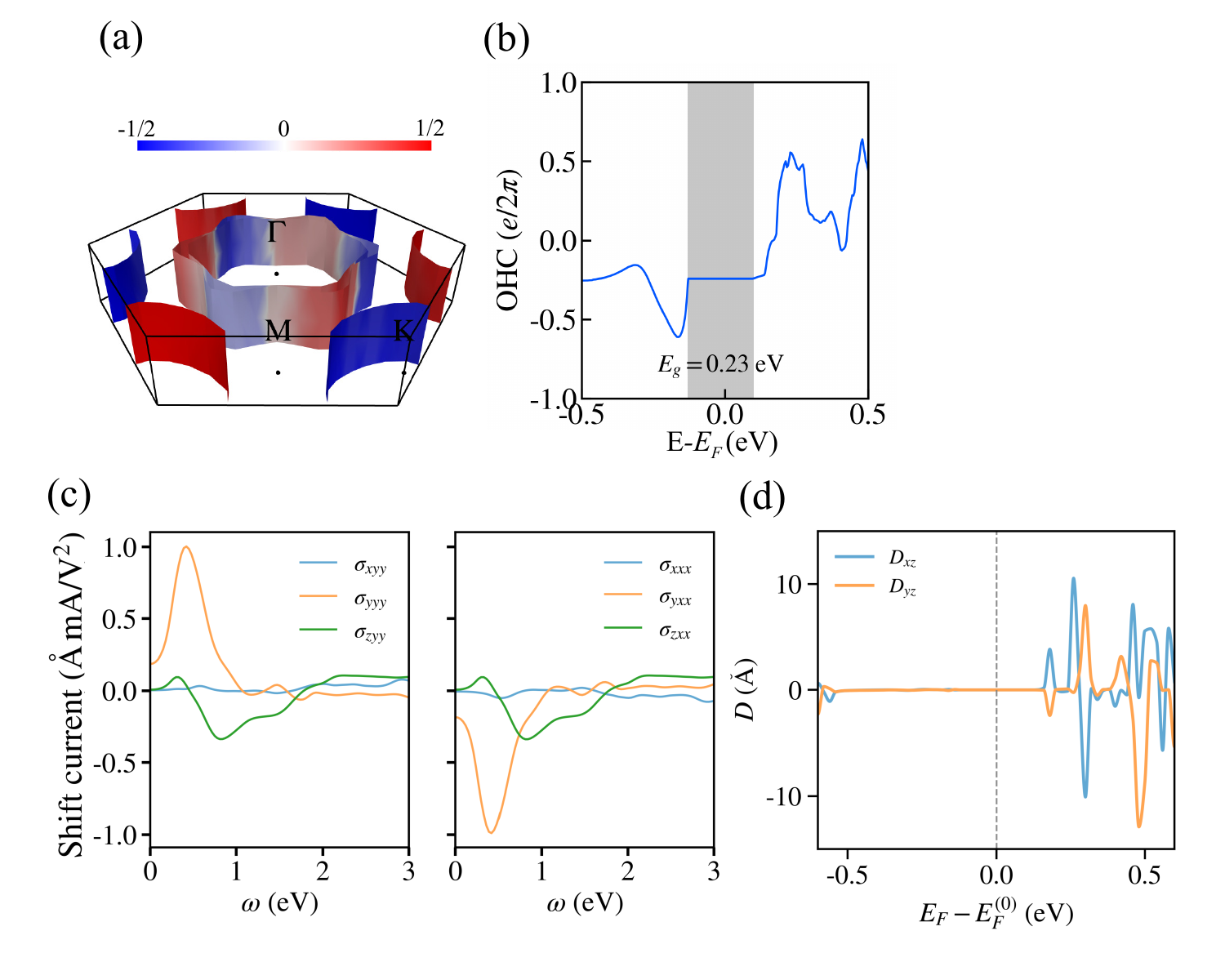}
 \caption{\textbf{Spin--valley locking and Berry-phase responses of Ta$_2$CS$_2$ and CrBr$_3$ hetrostructure.}
(a) Fermi surface colored by the out-of-plane spin expectation value $\langle S_z\rangle$ (color scale $\pm 1/2$), highlighting valley-contrasting spin polarization at the $K/K'$ valleys.
(b) Energy-dependent orbital Hall conductivity $\sigma^{L_z}_{yx}/(2\pi)$ versus $E-E_F$, with the bulk gap ($E_g=0.23$~eV) shaded.
(c) Frequency-dependent shift-current tensor components $\sigma_{abc}(\omega)$.
(d) Berry-curvature-dipole components $D_{xz}$ and $D_{yz}$ as a function of $E-E_F$ (dashed line marks $E_F$).}
   \label{fig:S9}
\end{figure*}

\begin{figure*}[t]
  \centering
  % try larger trims; units = left bottom right top
  \includegraphics[
    width=\textwidth,
    height=0.9\textheight,
    keepaspectratio,
    trim=0.0cm 4.0cm 0.0cm 0.5cm,
    clip
  ]{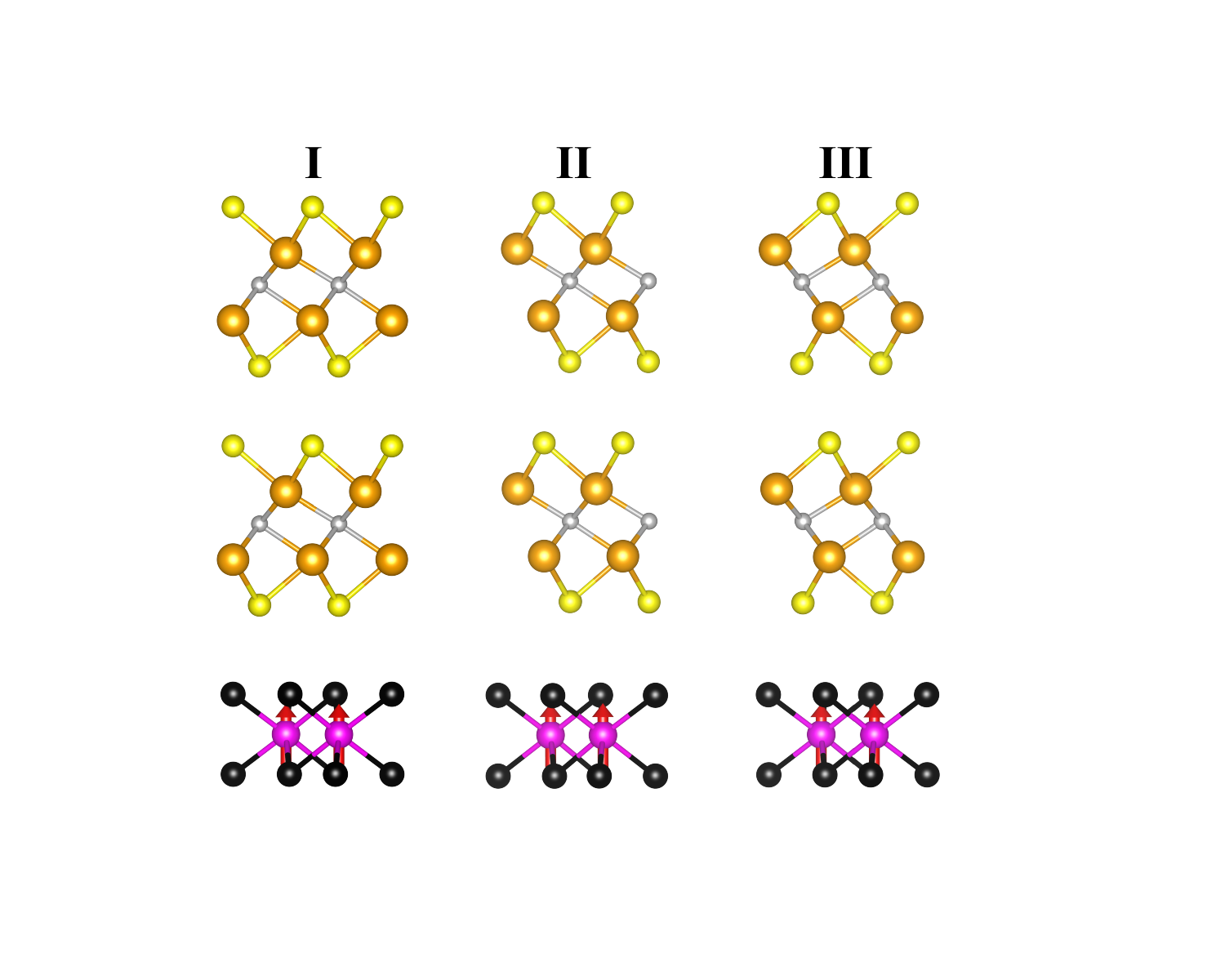}
 \caption{\textbf{Stacking configurations of bilayer M$_2$CS$_2$/CrBr$_3$ heterostructure.}
Side views of the three representative lateral configrations (I--III) considered in this work. Orange/gray/yellow spheres denote M/C/S atoms in M$_2$CS$_2$, and magenta/black spheres denote Cr/Br atoms in CrBr$_3$.}
   \label{fig:S10}
 \end{figure*}

\begin{figure*}[t]
  \centering
  % try larger trims; units = left bottom right top
  \includegraphics[
    width=\textwidth,
    height=0.9\textheight,
    keepaspectratio,
    trim=1.0cm 3.5cm 0.0cm 0.5cm,
    clip
  ]{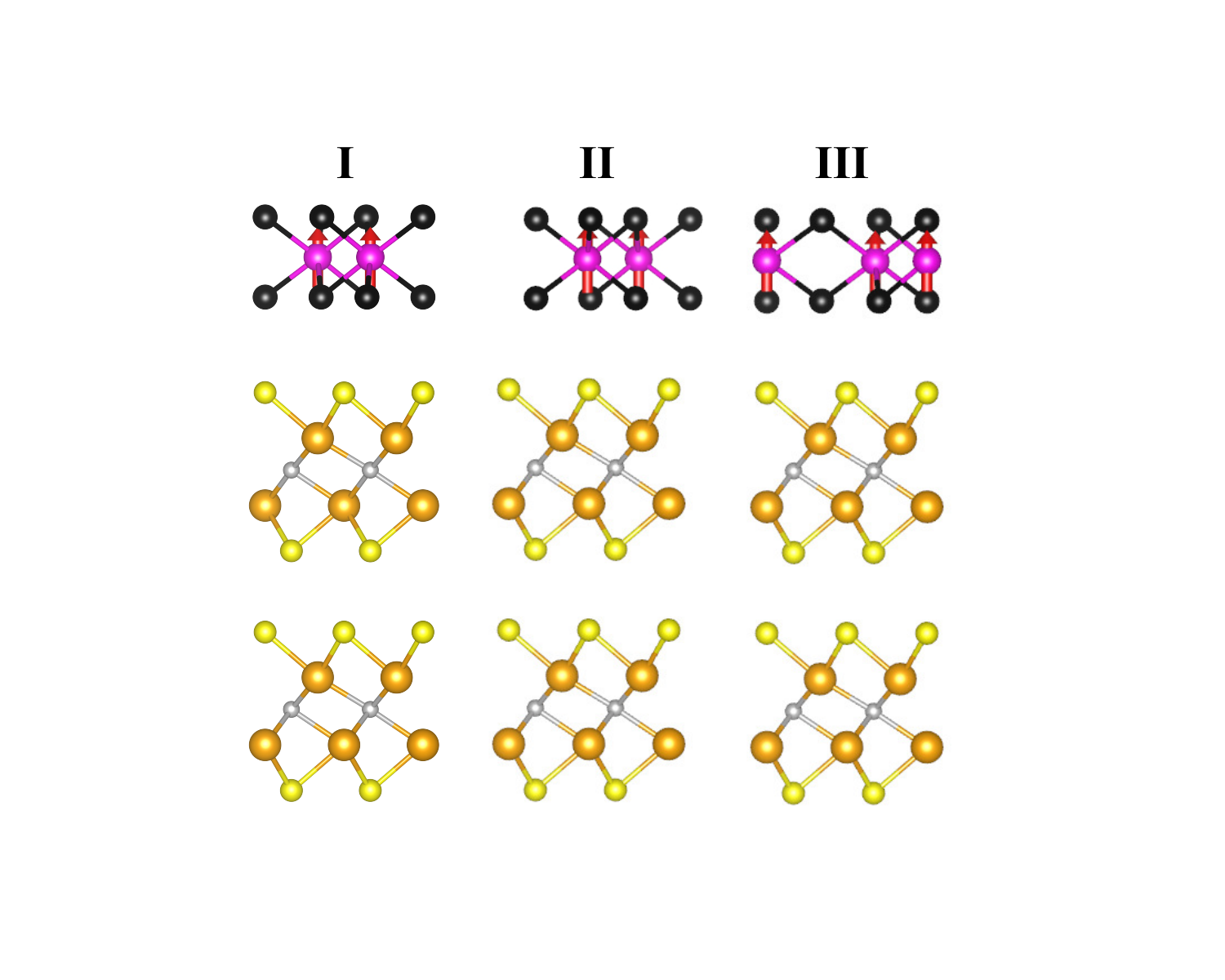}
 \caption{\textbf{Stacking configurations of bilayer CrBr$_3$/M$_2$CS$_2$ heterostructure.}
Side views of the three representative lateral configrations (I--III) considered in this work. Orange/gray/yellow spheres denote M/C/S atoms in M$_2$CS$_2$, and magenta/black spheres denote Cr/Br atoms in CrBr$_3$.}
   \label{fig:S11}
 \end{figure*}

\begin{table}[H]
\caption{Relative energies (meV) of different stacking configurations of bilayer M$_2$CS$_2$\allowbreak{}/CrBr$_3$ heterostructures, referenced to the most stable configuration in each case.}
\label{tab:relE_stacking_bilayer}
\centering
\begin{tabular}{l c c c}
\toprule
\multirow{2}{*}{System} & \multirow{2}{*}{Config.} & \multicolumn{2}{c}{$\Delta E$ (meV)} \\
\cmidrule(lr){3-4}
& & {CrBr$_3$ bottom} & {CrBr$_3$ top} \\
\midrule
\multirow{3}{*}{Ta$_2$CS$_2$}
& I   & 0    & 72.11 \\
& II  & 6.48  & 12.32 \\
& III & 69.02 & 0     \\
\midrule
\multirow{3}{*}{Nb$_2$CS$_2$}
& I   & 0     & 44.89 \\
& II  & 14.56 & 0     \\
& III & 48.10 & 3.12  \\
\bottomrule
\end{tabular}
\end{table}

\begin{table*}
\caption{Vacuum-referenced work function $\Phi = V_{\rm vac} - E_F$ 
for the three Ta$_2$CS$_2$@CrBr$_3$ stacking configurations.}
\label{tab:work_function}
\centering
\begin{tabular}{lccc}
\toprule
Configuration & $V_{\rm vac}$ (eV) & $E_F$ (eV) & $\Phi$ (eV) \\
\midrule
Bottom-stacked & 4.847 & $-0.404$ & 5.250 \\
Top-stacked (stable) & 4.866 & $-0.448$ & 5.314 \\
Top-stacked (metastable) & 4.865 & $-0.457$ & 5.322 \\
\bottomrule
\end{tabular}
\end{table*}

\begin{figure*}[t]
  \centering
  % try larger trims; units = left bottom right top
  \includegraphics[
    width=\textwidth,
    height=0.9\textheight,
    keepaspectratio,
    trim=0.0cm 0.7cm 0.0cm 0.5cm,
    clip
  ]{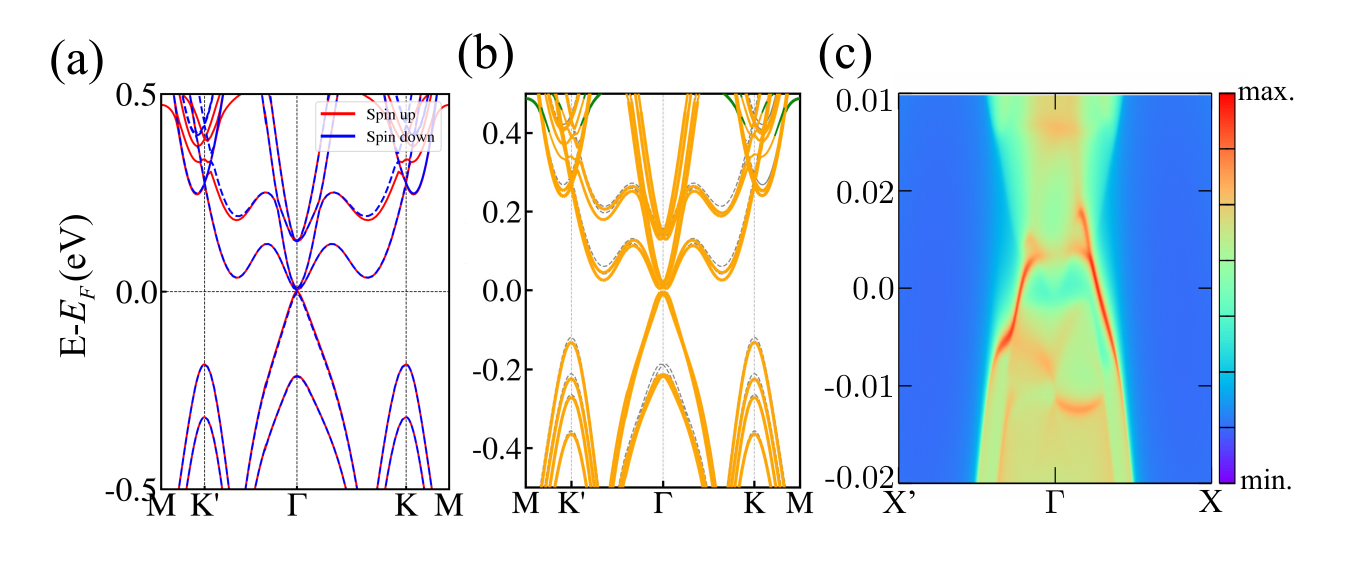}
\caption{\textbf{Quantum spin Hall state in bilayer Nb$_2$CS$_2$/CrBr$_3$ hetrostructure}. (a) Band structure without SOC and (b) with SOC. (b) Edge spectral function/band structure of a semi-infinite ribbon, showing gapless helical edge modes connecting valence and conduction continua.}
   \label{fig:S12}
 \end{figure*}

\clearpage

\begin{figure*}[t]
  \centering
  \includegraphics[width=\textwidth,trim=6pt 6pt 6pt 6pt,clip]{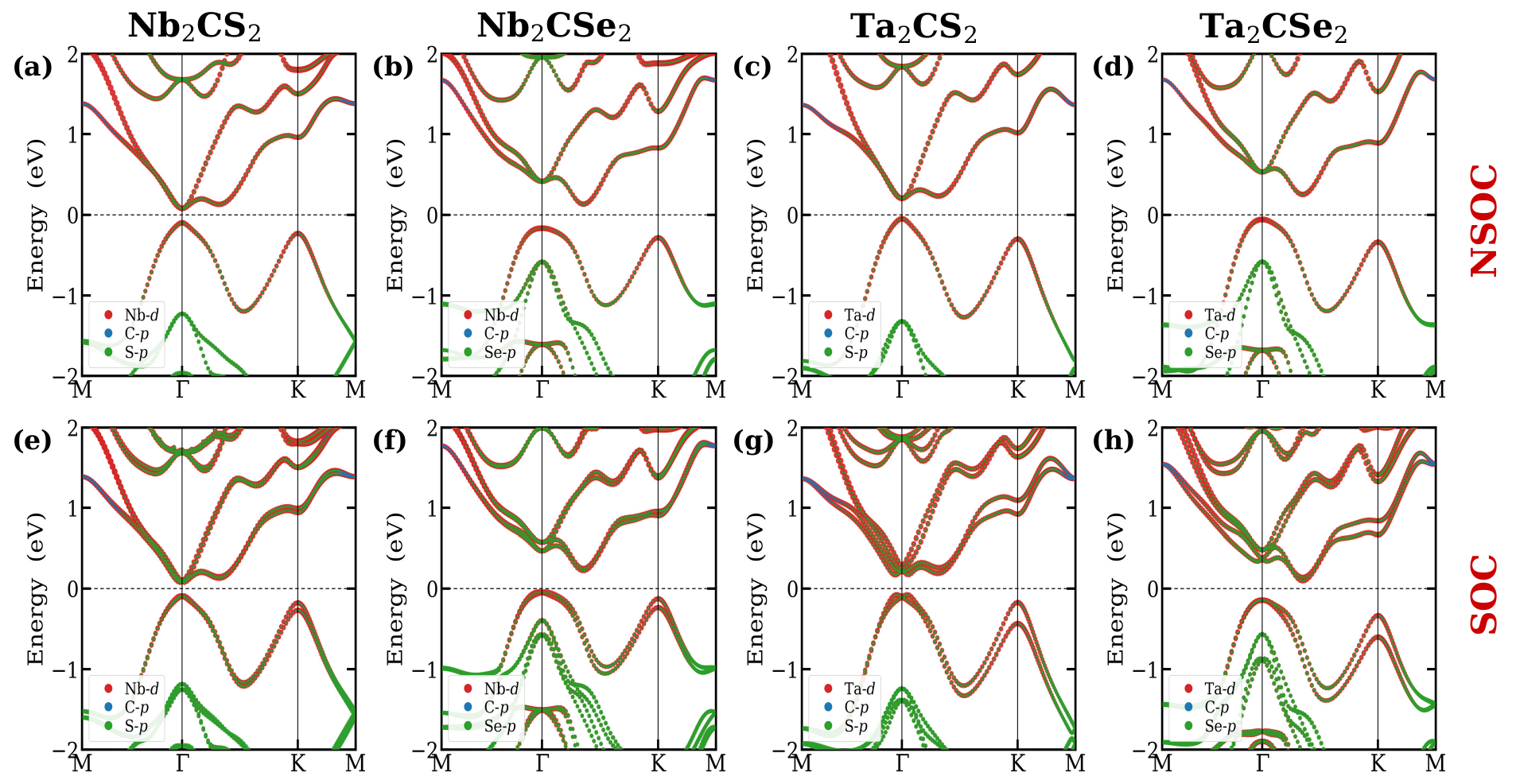}
    \caption{\textbf{Orbital-projected electronic band structures of monolayer Nb$_2$CS$_2$, Nb$_2$CSe$_2$, Ta$_2$CS$_2$, and Ta$_2$CSe$_2$}. Panels (a)–(d) show the non-spin-orbit-coupled (NSOC) results, while panels (e)–(h) show the corresponding spin-orbit-coupled (SOC) band structures. The projected weights are represented by colored markers, with red, blue, and green denoting transition-metal $d$, C-$p$, and chalcogen-$p$ states, respectively. The horizontal dashed line marks the Fermi level set to 0 eV. The SOC results show the modification of the near-Fermi-level dispersions and the orbital character compared to the NSOC case.}
    \label{fig:fig13}
\end{figure*}

\clearpage

\begin{figure*}[t]
  \centering
  \includegraphics[width=\textwidth,trim=6pt 6pt 6pt 6pt,clip]{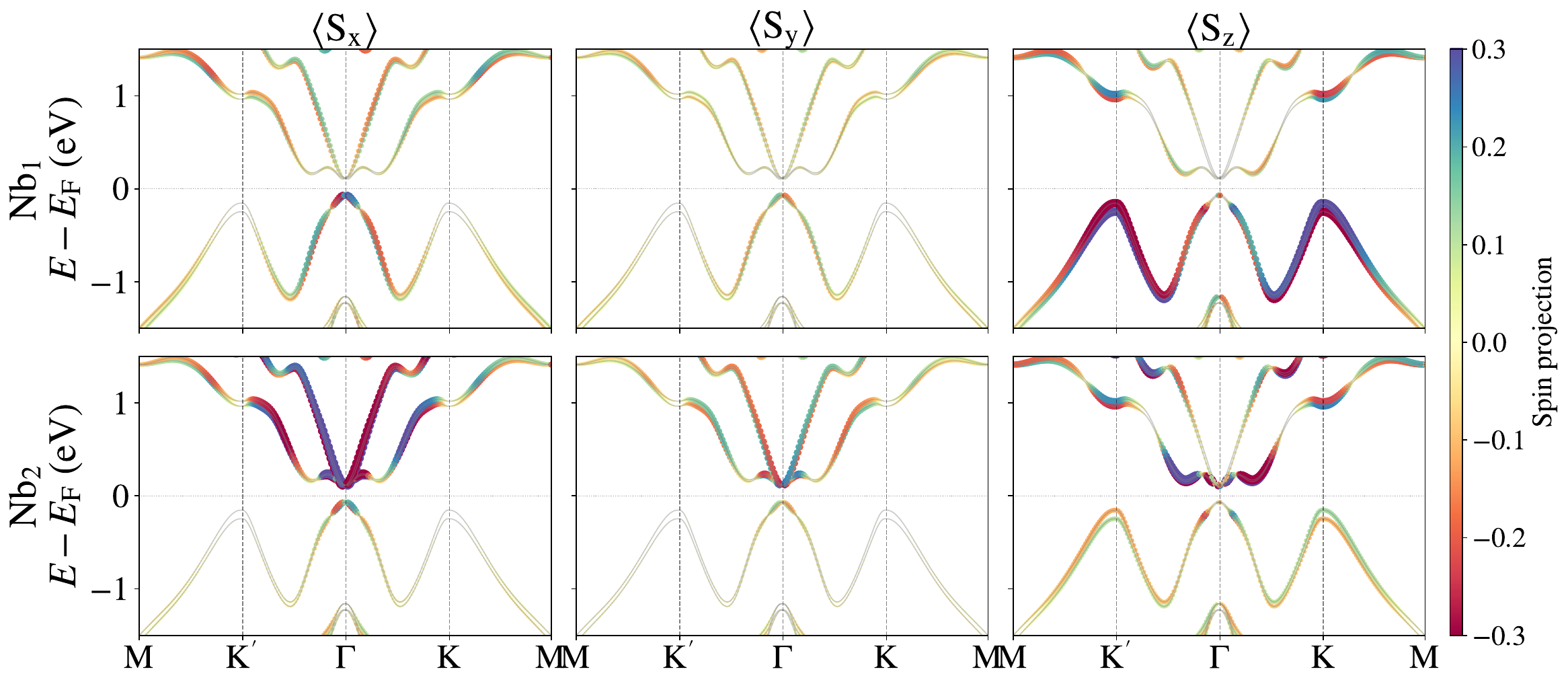}
    \caption{Atom-resolved spin projections for monolayer Nb$_2$CS$_2$ 
along the M--K$'$--$\Gamma$--K--M path. Rows correspond to 
the two inequivalent Nb sublattices: Nb$_1$ at the lower $z$ 
position (top) and Nb$_2$ at the upper $z$ position (bottom). Nb$_1$ dominates 
the valence-band region, including the $d_{z^2}$ states near 
$\Gamma$ and the deeper $L_z=\mp 2$ states at K/K$'$, while 
Nb$_2$ dominates the conduction-band region, mirroring the 
sublattice segregation discussed in the main text for monolayer 
Ta$_2$CS$_2$.}
    \label{fig:fig14}
\end{figure*}

\begin{figure*}[t]
  \centering
  \includegraphics[width=\textwidth,trim=6pt 6pt 6pt 6pt,clip]{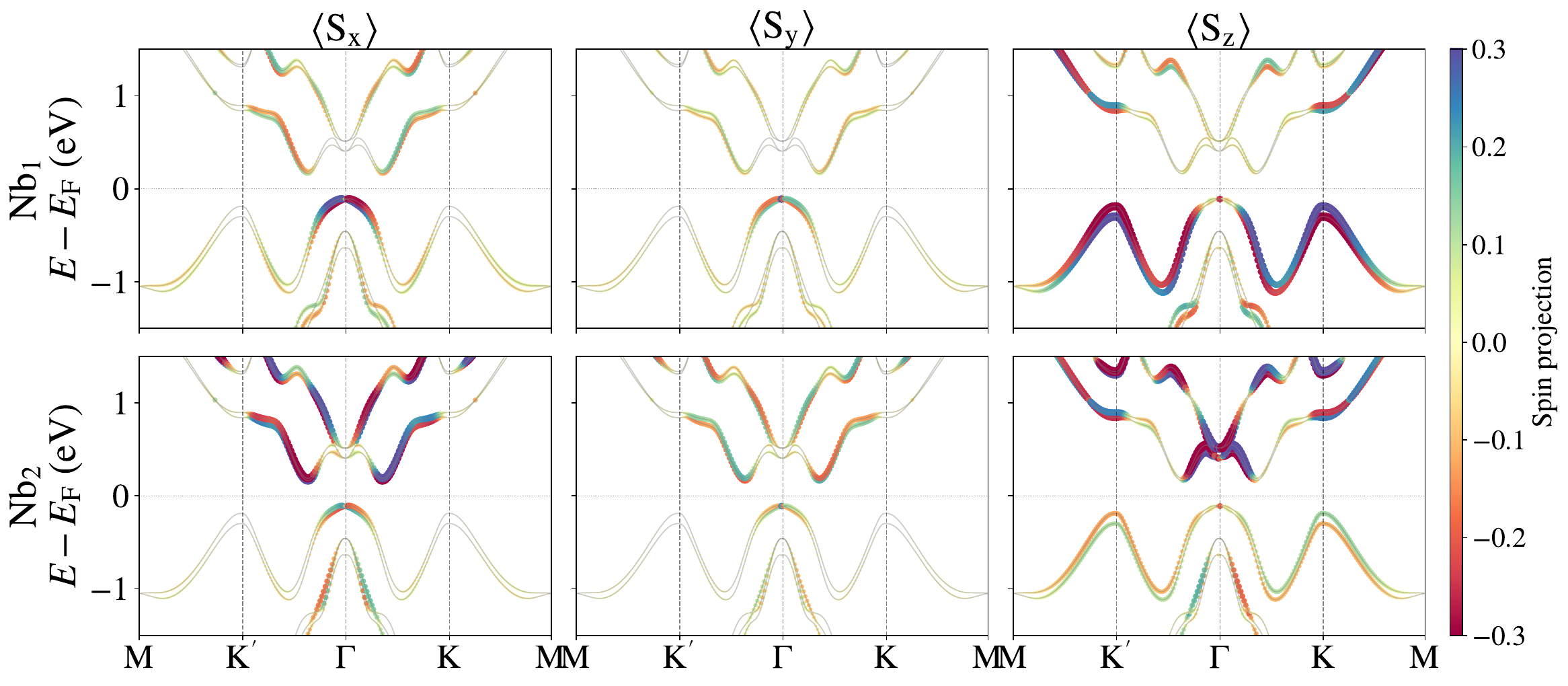}
    \caption{Atom-resolved spin projections for monolayer Nb$_2$CSe$_2$ 
along the M--K$'$--$\Gamma$--K--M path. Rows correspond to 
the two inequivalent Nb sublattices: Nb$_1$ at the lower $z$ 
position (top) and Nb$_2$ at the upper $z$ position (bottom). Nb$_1$ dominates 
the valence-band region, including the $d_{z^2}$ states near 
$\Gamma$ and the deeper $L_z=\mp 2$ states at K/K$'$, while 
Nb$_2$ dominates the conduction-band region.}
    \label{fig:fig15}
\end{figure*}

\begin{figure*}[t]
  \centering
  \includegraphics[width=\textwidth,trim=6pt 6pt 6pt 6pt,clip]{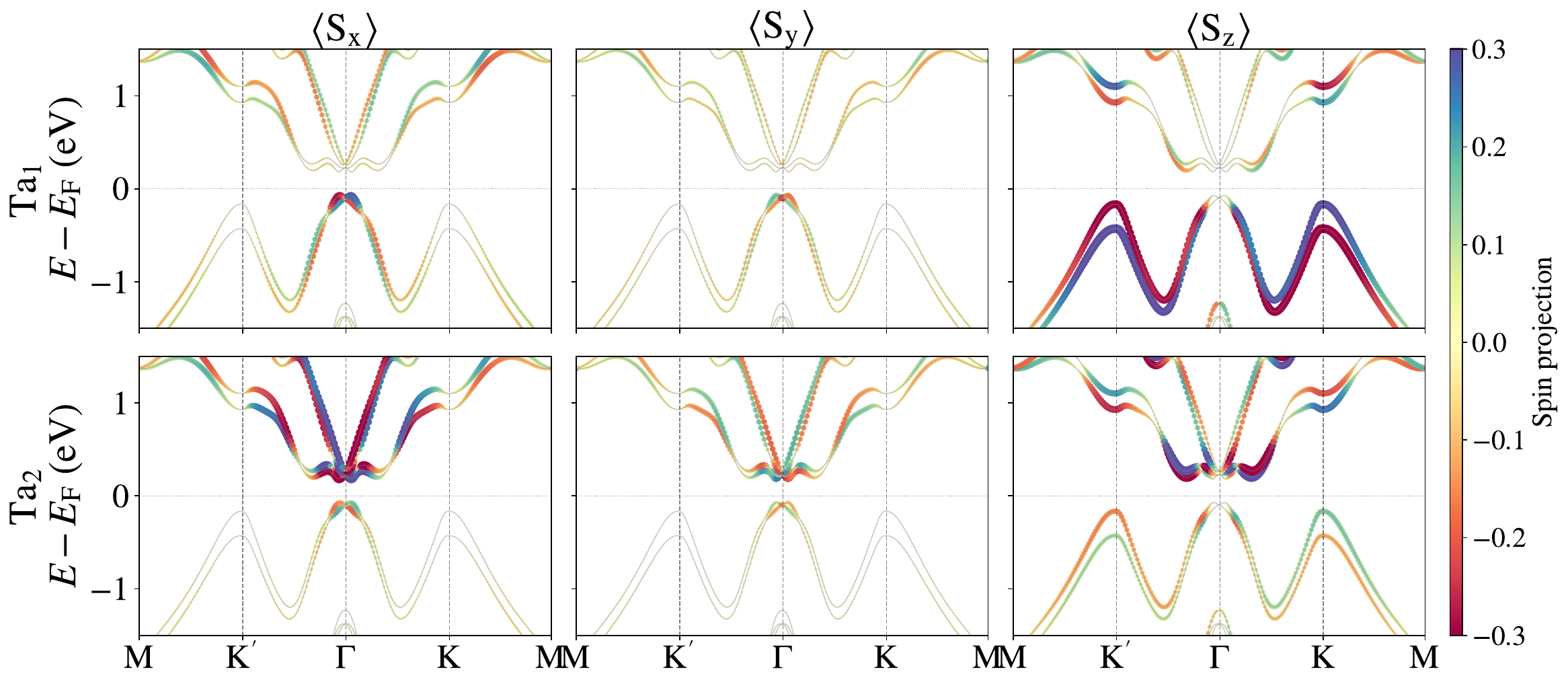}
    \caption{Atom-resolved spin projections for monolayer Ta$_2$CS$_2$ 
along the M--K$'$--$\Gamma$--K--M path. Rows correspond to 
the two inequivalent Nb sublattices: Nb$_1$ at the lower $z$ 
position (top) and Nb$_2$ at the upper $z$ position (bottom). Nb$_1$ dominates 
the valence-band region, including the $d_{z^2}$ states near 
$\Gamma$ and the deeper $L_z=\mp 2$ states at K/K$'$, while 
Nb$_2$ dominates the conduction-band region.}
    \label{fig:fig16}
\end{figure*}

\begin{figure*}[t]
  \centering
  \includegraphics[width=\textwidth,trim=6pt 6pt 6pt 6pt,clip]{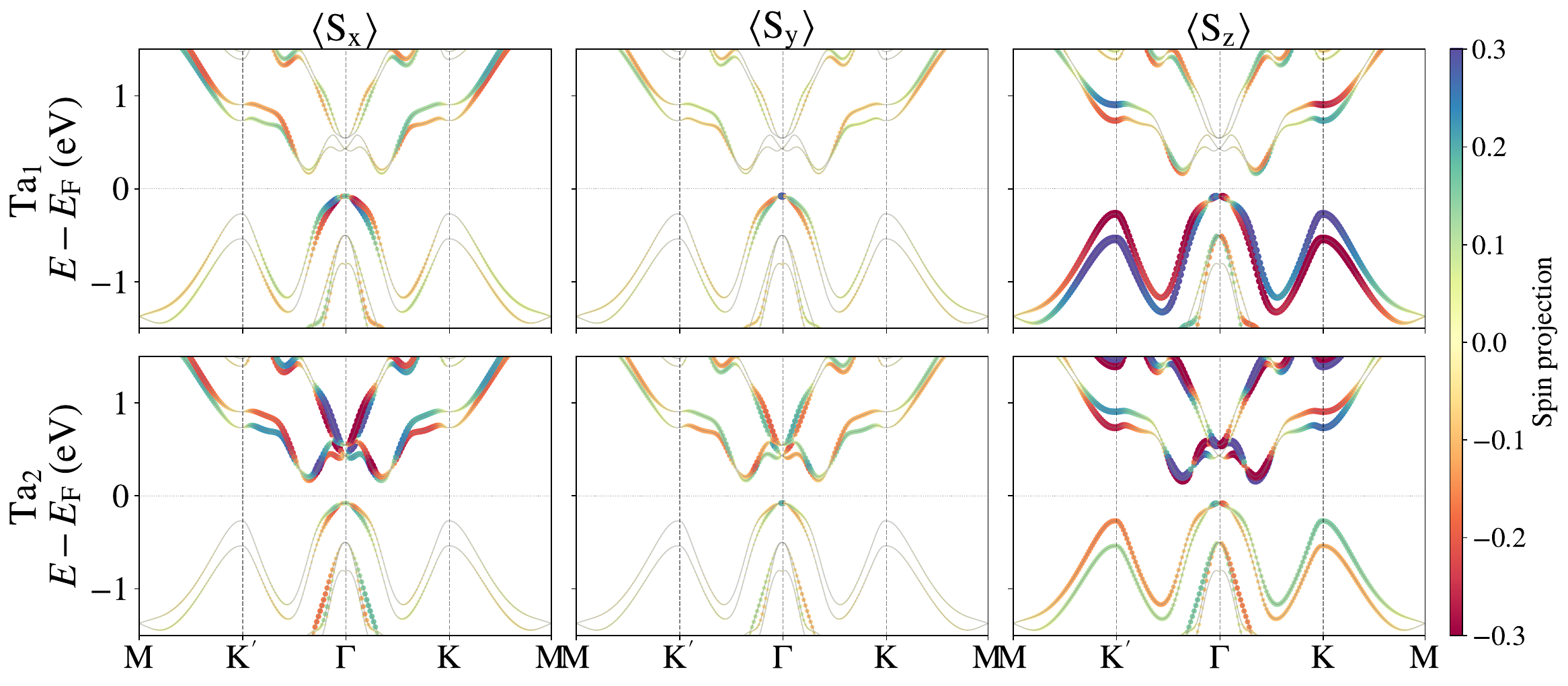}
    \caption{Atom-resolved spin projections for monolayer Ta$_2$CSe$_2$ 
along the M--K$'$--$\Gamma$--K--M path. Rows correspond to 
the two inequivalent Nb sublattices: Nb$_1$ at the lower $z$ 
position (top) and Nb$_2$ at the upper $z$ position (bottom). Nb$_1$ dominates 
the valence-band region, including the $d_{z^2}$ states near 
$\Gamma$ and the deeper $L_z=\mp 2$ states at K/K$'$, while 
Nb$_2$ dominates the conduction-band region.}
    \label{fig:fig17}
\end{figure*}

\begin{figure*}[t]
  \centering
  \includegraphics[width=\textwidth,trim=6pt 6pt 6pt 6pt,clip]{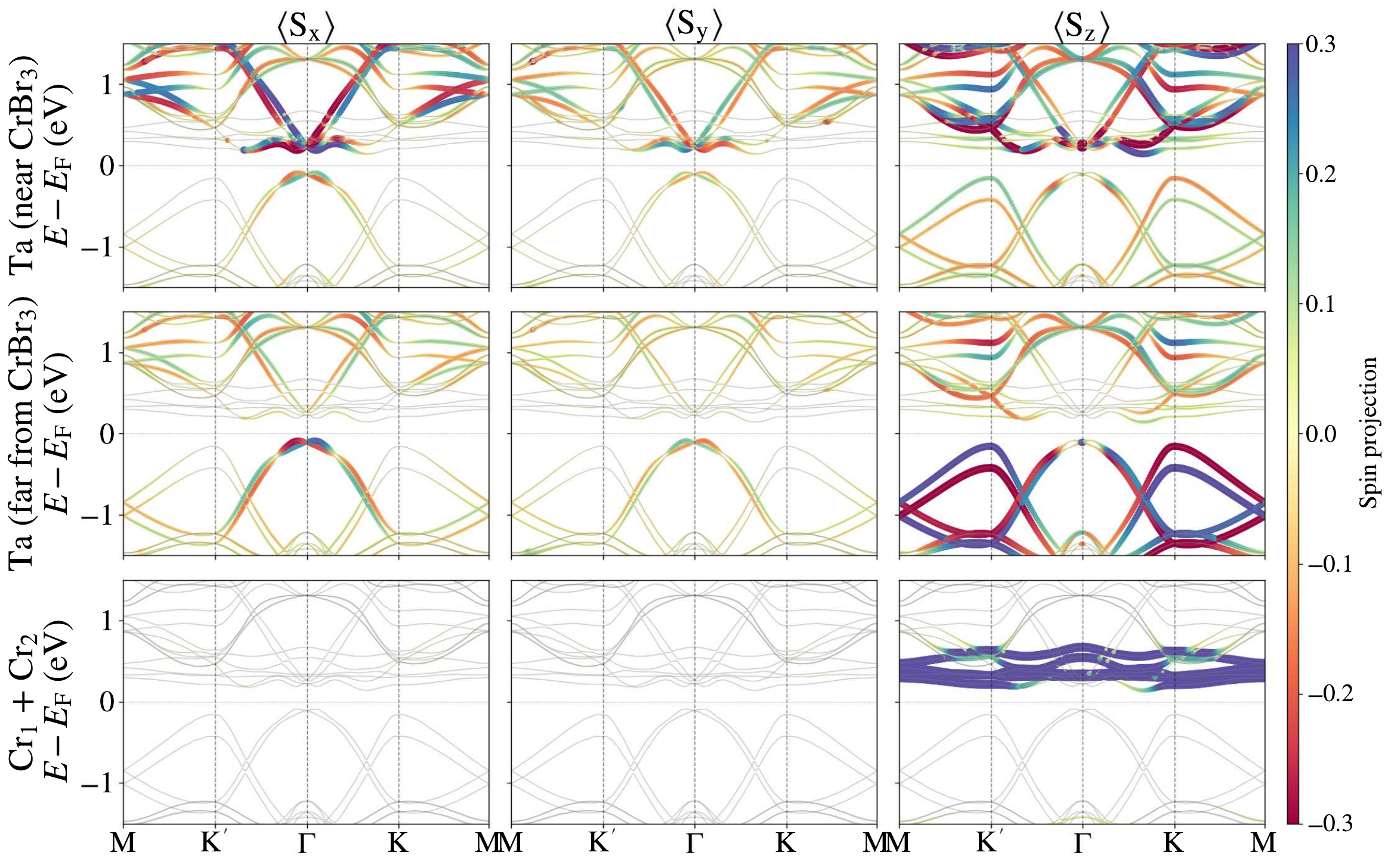}
\caption{Ion- and spin-resolved band structure of Ta$_2$CS$_2$@CrBr$_3$ 
(CrBr$_3$ on bottom) projected onto the total $d$-manifold. Each panel 
shows the band dispersion along $M$--$K'$--$\Gamma$--$K$--$M$, with 
marker size and color indicating the magnitude and sign of the spin 
projection $\langle S_i \rangle$ ($i = x, y, z$) summed over all five 
$d$ orbitals. Top row: Ta layer proximal to CrBr$_3$; middle row: distal 
Ta layer; bottom row: Cr$_1$ + Cr$_2$ combined. The proximity-induced 
exchange interaction from CrBr$_3$ predominantly affects the Ta layer 
closest to the substrate, introducing a momentum-dependent anisotropy 
in the already Rashba-split conduction bands: the CB edge along 
$\Gamma$--$K$ is renormalized downward relative to $\Gamma$--$K'$, 
generating a energy asymmetry of $\sim$50~meV. The Cr $d$ states 
appear as flat, nearly dispersionless bands near $E_F$ with dominant 
$\langle S_z \rangle$ character, consistent with the out-of-plane 
ferromagnetic ordering of CrBr$_3$.}
    \label{fig:fig18}
\end{figure*}

\begin{figure*}[t]
  \centering
  \includegraphics[width=\textwidth,trim=6pt 6pt 6pt 6pt,clip]{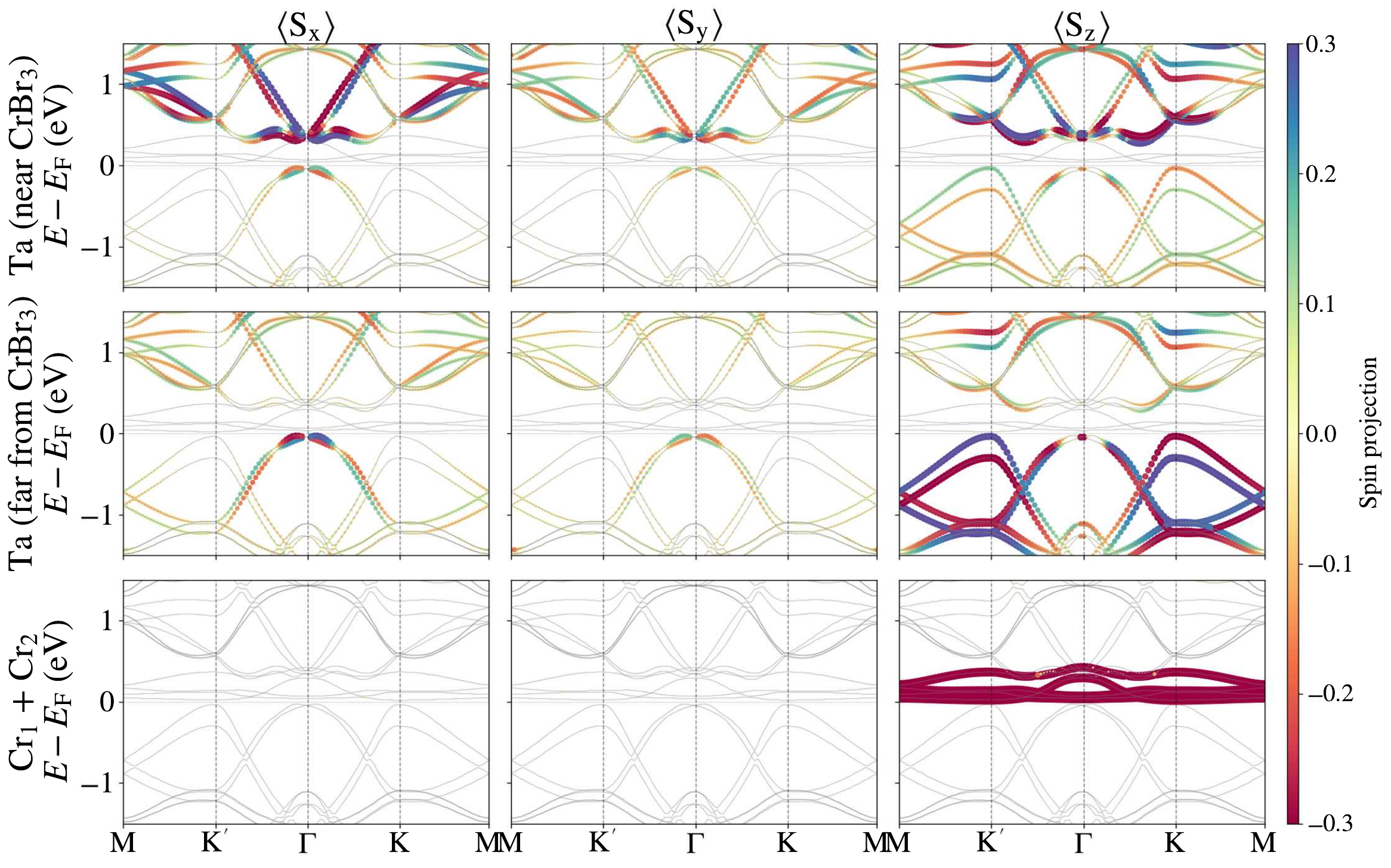}
    \caption{Ion- and spin-resolved band structure of CrBr$_3$@Ta$_2$CS$_2$ 
(CrBr$_3$ on top) projected onto the total $d$-manifold. The layout and 
color convention are identical to Fig.~S18. Compared to the bottom-substrate 
configuration, flipping CrBr$_3$ to the top reverses which Ta layer 
experiences the proximity-induced exchange interaction, now acting on the 
upper Ta layer. Most notably, the Cr $d$ states (bottom row) shift 
downward in energy relative to the bottom-substrate case, intruding into 
the conduction band region and substantially reducing the effective band 
gap. The dominant spin character of the affected conduction band states 
is reversed compared to Fig.~S18, consistent with the exchange field now 
coupling to the opposite Ta sublayer. This substrate-position dependence 
demonstrates that the proximity effect in Ta$_2$CS$_2$@CrBr$_3$ is 
highly sensitive to the interface geometry, offering a structural handle 
to tune both the band gap and the spin polarization of the conduction 
bands.}
    \label{fig:fig19}
\end{figure*}

\begin{figure*}[t]
  \centering
  \includegraphics[width=\textwidth,trim=6pt 6pt 6pt 6pt,clip]{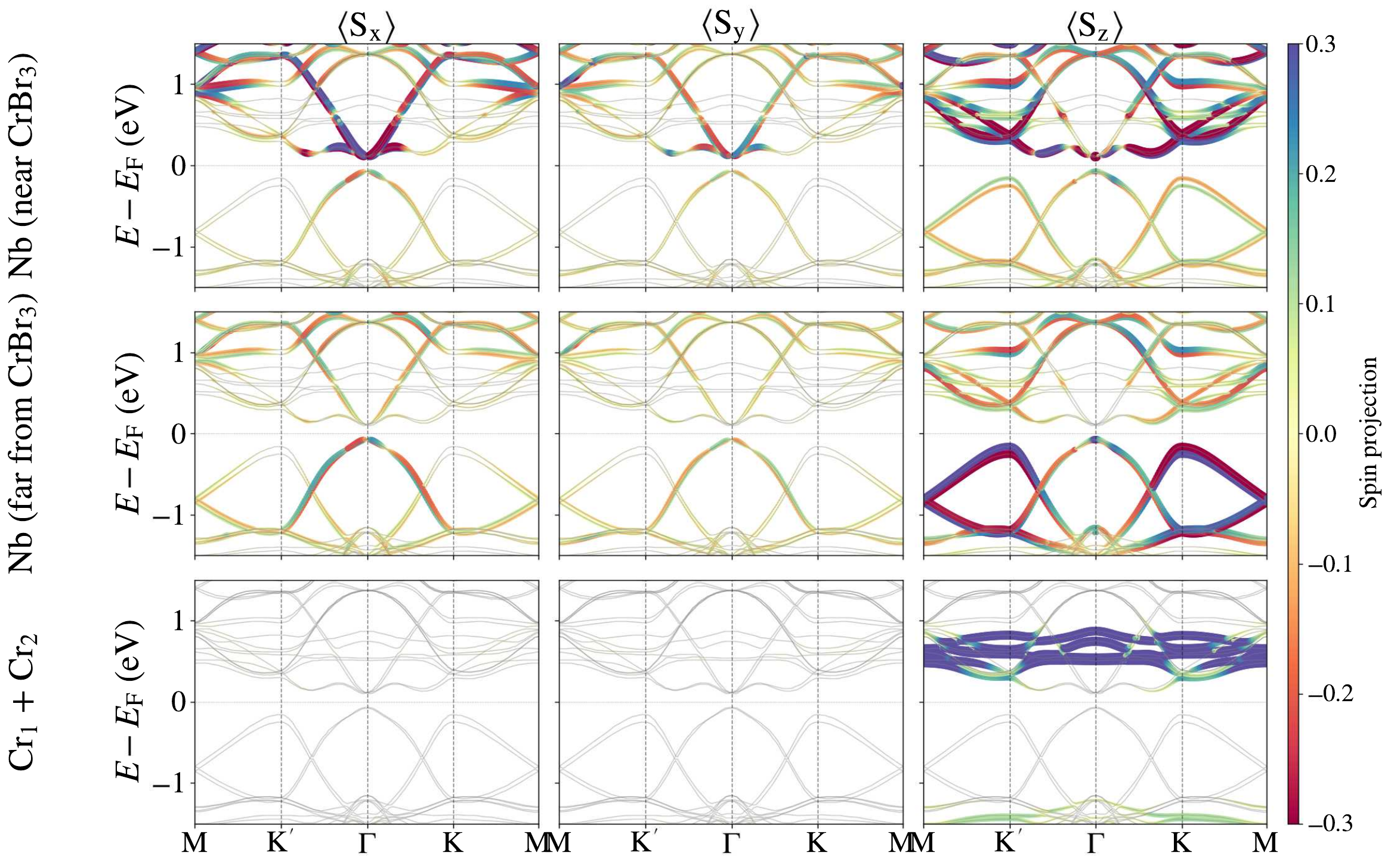}
   \caption{Ion- and spin-resolved band structure of Nb$_2$CS$_2$@CrBr$_3$ 
(CrBr$_3$ on bottom) projected onto the total $d$-manifold. Each panel 
shows the band dispersion along $M$--$K'$--$\Gamma$--$K$--$M$, with 
marker size and color indicating the magnitude and sign of the spin 
projection $\langle S_i \rangle$ ($i = x, y, z$) summed over all five 
$d$ orbitals. Top row: Nb layer proximal to CrBr$_3$; middle row: 
distal Nb layer; bottom row: Cr$_1$ + Cr$_2$ combined. As in the 
Ta$_2$CS$_2$ case, the proximity-induced exchange interaction 
selectively perturbs the nearest Nb layer, introducing a 
momentum-dependent anisotropy in the Rashba-split conduction bands 
with an asymmetric renormalization along $\Gamma$--$K$ versus 
$\Gamma$--$K'$. The Cr $d$ states are nearly dispersionless near 
$E_F$ with dominant $\langle S_z \rangle$ character.}
    \label{fig:figS20}
\end{figure*}
\clearpage

\begin{figure*}[t]
  \centering
  \includegraphics[width=\textwidth,trim=6pt 6pt 6pt 6pt,clip]{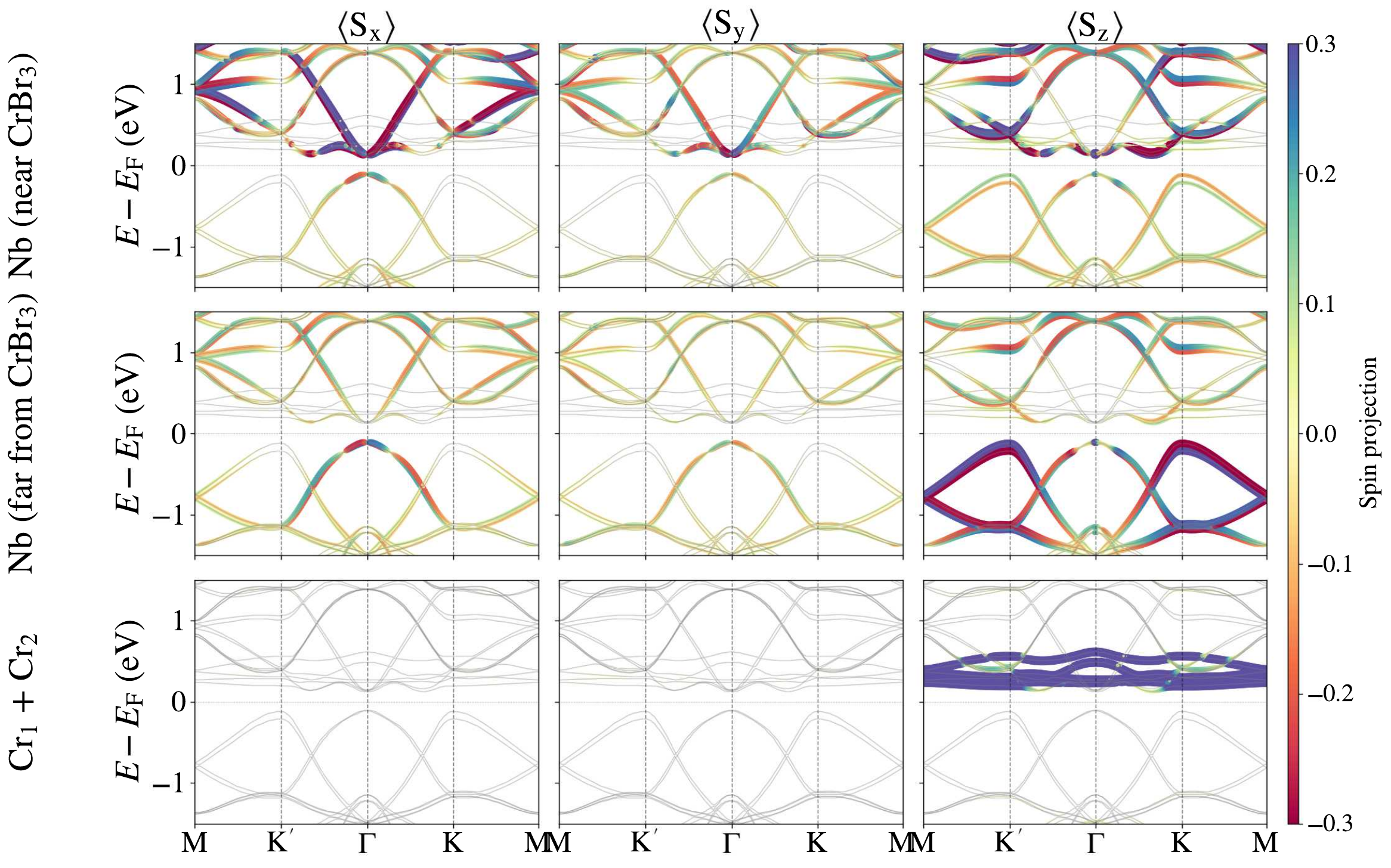}
   \caption{Same as Fig.~\ref{fig:figS20} but with CrBr$_3$ placed on top of 
Nb$_2$CS$_2$.}
    \label{fig:figS21}
\end{figure*}
\clearpage

\begin{figure*}[t]
  \centering
  \includegraphics[width=\textwidth,trim=6pt 6pt 6pt 6pt,clip]{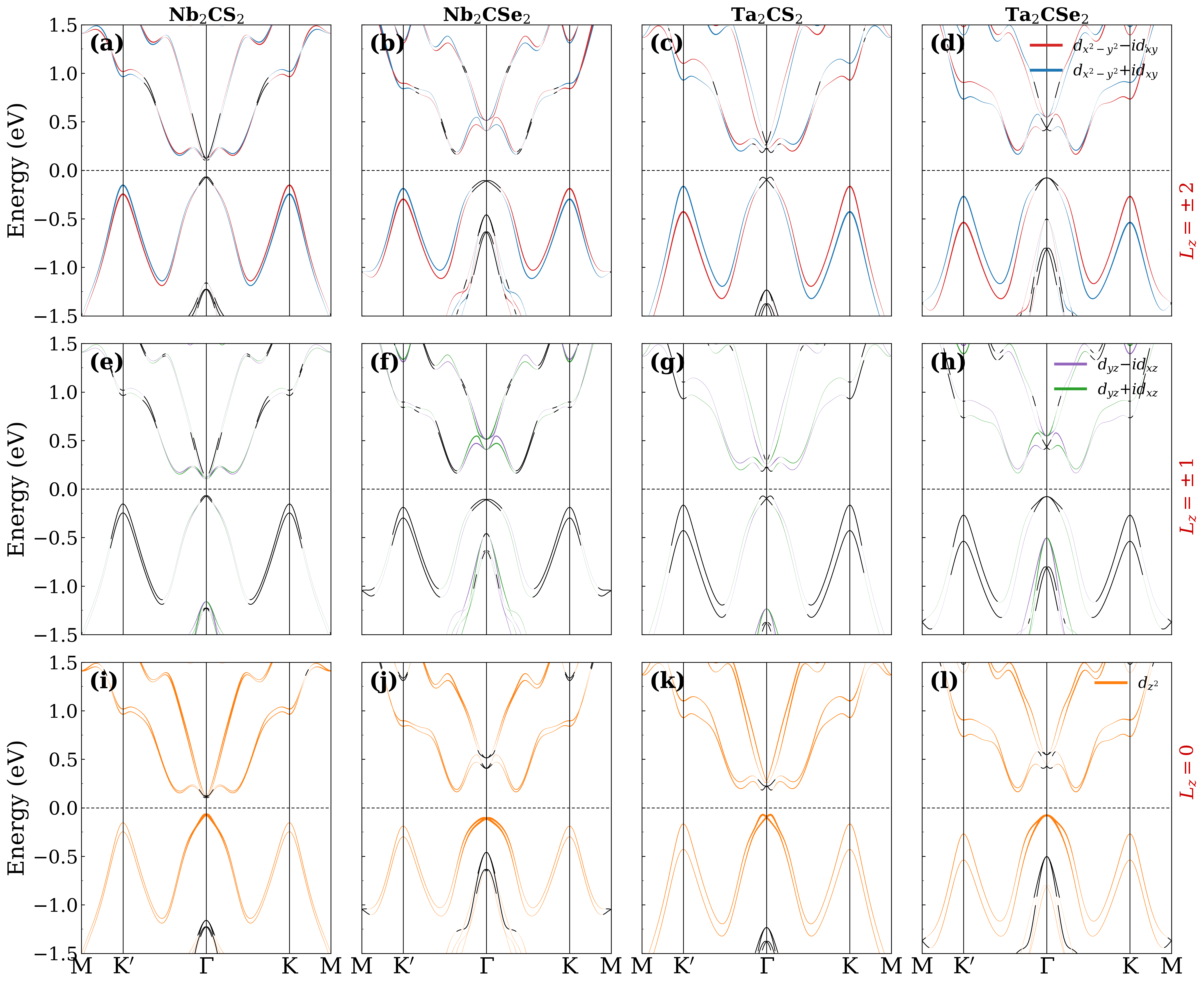}
    \caption{Orbital angular-momentum projections of the SOC band 
structures of monolayer M$_2$CX$_2$ (M = Nb, Ta; X = S, Se) along 
the $M$--$K'$--$\Gamma$--$K$--$M$ path. Columns correspond to the 
four compounds: (a,e,i) Nb$_2$CS$_2$, (b,f,j) Nb$_2$CSe$_2$, 
(c,g,k) Ta$_2$CS$_2$, and (d,h,l) Ta$_2$CSe$_2$. Rows correspond 
to the three angular-momentum eigenstate manifolds: 
(a--d) $L_z=\pm 2$ ($d_{x^2-y^2}\mp i d_{xy}$, red/blue for 
$L_z=-2/+2$); (e--h) $L_z=\pm 1$ ($d_{yz}\mp i d_{xz}$, 
purple/green for $L_z=-1/+1$); (i--l) $L_z=0$ ($d_{z^2}$, 
orange). The horizontal dashed line marks $E_F$. In all four 
compounds, the $\Gamma$-point band edges are dominated by 
$L_z=0$ ($d_{z^2}$) character, whereas the off-$\Gamma$ valley 
regions of both VBM and CBM are dominated by $L_z=\pm 2$ 
character. The $L_z=\pm 1$ manifold contributes mainly to the 
deeper bands away from the band edges.}
    \label{fig:figS22}
\end{figure*}

\begin{figure*}[t]
  \centering
  % try larger trims; units = left bottom right top
  \includegraphics[
    width=\textwidth,
    height=0.9\textheight,
    keepaspectratio,
    trim=0.1cm 1.2cm 0.7cm 0.0cm,
    clip
  ]{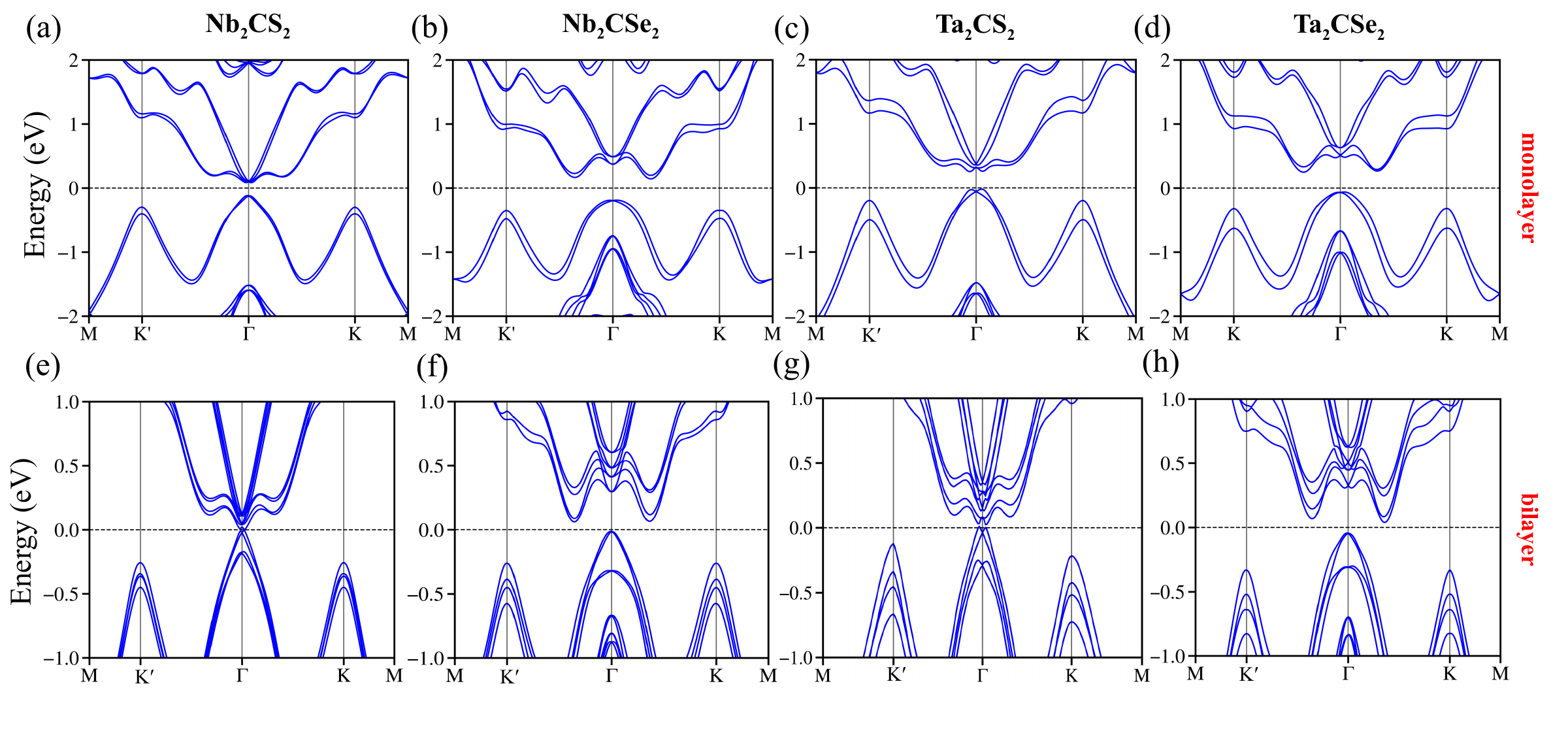}
  \caption{Band structures of monolayer M$_2$CX$_2$ (M = Nb, Ta; X = S, Se) calculated with the HSE06 hybrid functional. Upper row: without SOC. Lower row: with SOC.}
  \label{fig:figS23}
\end{figure*}

\end{document}

%% file: main.bbl
\providecommand{\latin}[1]{#1}
\makeatletter
\providecommand{\doi}
  {\begingroup\let\do\@makeother\dospecials
  \catcode`\{=1 \catcode`\}=2 \doi@aux}
\providecommand{\doi@aux}[1]{\endgroup\texttt{#1}}
\makeatother
\providecommand*\mcitethebibliography{\thebibliography}
\csname @ifundefined\endcsname{endmcitethebibliography}  {\let\endmcitethebibliography\endthebibliography}{}